\documentclass{jfm}
\usepackage{amsmath}
\usepackage{newtxtext}
\usepackage{newtxmath}
\usepackage{mathrsfs}
\usepackage{xcolor}
\usepackage{graphicx}
\usepackage{bm}
\usepackage{natbib}
\usepackage{epstopdf}
\usepackage{enumitem}
\usepackage{comment}
\usepackage{tikz}
\usetikzlibrary{decorations.pathreplacing}
\usepackage{hyperref}
\hypersetup{colorlinks=true,
            linkcolor=red,
            anchorcolor=blue,
            citecolor=blue}

\newcommand{\ri}{\mathop{\rm i}\nolimits}
\newcommand{\re}{\mathop{\rm e}\nolimits}

\title{
Excitation of non-modal perturbations in hypersonic boundary layers by free stream forcing. Part II: asymptotic theory and key mechanisms
}
\author[M. Dong, M. Sun, Q. Song and L. Zhao]%
{ Ming\ns DONG$^{1}$\thanks{Email address for correspondence: dongming@imech.ac.cn},
 Mingze\ns SUN$^{2}$,
 Qinyang\ns SONG$^{2}$
and Lei\ns ZHAO$^{2,3}$
  \thanks{Email address for correspondence: lei\_zhao@tju.edu.cn}}
\affiliation{$^1$State Key Laboratory of Nonlinear Mechanics, Institute of Mechanics, Chinese Academy of Sciences, Beijing 100190, China\\
$^2$Department of Mechanics, Tianjin University, Tianjin, 300072, China\\
$^3$National Key Laboratory of Vehicle Power System, Tianjin 300350, PR China
}

\begin{document}
\maketitle

\begin{abstract}
Recently, Zhao \& Dong (J. Fluid Mech. 2025, vol. 1013: A44) developed a high-efficiency, high-accuracy numerical framework, the shock-fitting harmonic  {linearised} Navier–Stokes (SF-HLNS) approach, which enables a systematic study of the receptivity of non-modal perturbations in hypersonic blunt-body boundary layers over a wide parameter range. In this Part II, we employ a high-Reynolds-number asymptotic analysis to elucidate the physical mechanism of the receptivity process. A distinct slow-down convection mechanism is identified in the nose region, amplifying the perturbation streamwise vorticity from the post-shock position to the  boundary layer around the stagnation point by a factor of $O(R^{1/2})$, where $R$ is the Reynolds number based on nose radius. Downstream, the lift-up mechanism further leads to a transient growth of the perturbation streamwise velocity up to {\color{red}an amplitude of} $O(R)$. Based on these {\color{red}mechanisms}, a reduced model is developed to predict the downstream evolution of the non-modal perturbations initiated by receptivity, whose predictions agree well with SF-HLNS calculations.  This model can also be used to investigate the effects of wall temperature and nose radius on non‑modal receptivity efficiency, as will be detailed in {Part} III of this work series.
\end{abstract}



\section{Introduction}
In the design of high-speed vehicles, accurate prediction of laminar–turbulent transition remains challenging due to its high sensitivity to numerous influencing factors. {\color{red}Transition routes can be broadly classified into two categories: the natural route, wherein exponential growth of modal instabilities dominates the early laminar phase, and the bypass route, wherein this exponential growth stage is bypassed.} It is well established that as freestream perturbation level intensifies from low to high, the dominant transition mechanism shifts progressively from natural to bypass \citep{Fedorov2011transition,Zhong2012direct}. Notably, freestream forcing is not the sole factor influencing transition; even under a fixed background noise level, increasing nose bluntness can still induce a distinct shift in transition mechanism, as documented experimentally \citep{stetson1983nosetip,lysenko1990influence,kosinov1990experiments,jewell2018transition,borovoy2022laminar}. Since moderate bluntness is often preferred for thermal protection, predicting bypass transition in moderate‑bluntness configurations becomes an essential problem. However, despite extensive studies on bypass transition mechanisms \citep{Durbin2007,Zhao2020}, models for accurately predicting bypass transition onset remain scarce, especially for hypersonic boundary layers. This challenge  arises from the inherent complexity of the bypass transition process.

In natural transition, the laminar phase is governed by the exponential amplification of modal instability modes, a process readily predicted by {the} linear stability theory (LST) under the parallel-flow assumption \citep{Mack1975linear}. {\color{red}For hypersonic boundary layers, multiple instability modes may emerge, among which the Mack second mode is typically the most unstable. However, because the parallel-flow assumption introduces quantitative errors at finite Reynolds numbers, the linear parabolised stability equations (LPSE) have been formulated to}   account for the weak non-parallelism, thereby providing a more accurate prediction of modal amplitude evolution \citep{Herbert1997parabolized}. {\color{red}Nevertherless,  the linear approaches remain insufficient to describe the perturbation evolution during the nonlinear phase.} Therefore, the nonlinear {parabolised} stability equations (NPSE) have been developed, which retain triad interactions among Fourier components, enabling the calculating  nonlinear saturation and harmonic generation \citep{songrj2023effect,songrj2024principle}.

In contrast, bypass transition is not governed by modal instabilities; rather, it proceeds sequentially through {\color{red}the following steps: the excitation of non-modal perturbations by freestream perturbations, their transient growth leading to the formation of longitudinal streaks \citep{Leib1999}, the subsequent nonlinear saturation and secondary instability\citep{ricco2011evolution}, and finally, the emergence of turbulent spots culminating in full breakdown  to turbulence \citep{Zhang2018}. The formation of the longitudinal streaks are confirmed in experiments for both incompressible \citep{Matsubara2001} and hypersonic \citep{Kennedy2022} configurations.} The non-modal perturbations consist of a series of modal components, which are not mutually orthogonal, allowing transient growth to occur even if all modal components are exponentially decaying. Detailed literature reviews of {\color{cyan}natural} and bypass transitions can be found in \cite{Dong2021asymptotic} and \cite{zhao2025excitation} ({hereafter denoted by ZD25}), respectively.

The formation of downstream non-modal perturbations can be physically attributed to the lift‑up mechanism \citep{Landahl1980,Brandt2014}. In this process,  roll‑type structures in the cross‑plane drive upward and downward fluid motion, shearing the base flow and  generating alternating fast‑ and slow‑speed streaks. 
A widely used method for computing non-modal evolution is the optimal growth theory (OGT), which identifies the maximum energy amplification between input and output positions via optimization, such as adjoint methods \citep{Andersson1999,paredes2019nonmodal,paredes2020mechanism}. However, OGT suffers from a critical limitation: it lacks a physical link to external forcing. {\color{red} Consequently, the optimal input structures identified by OGT may not correspond to physically realizable freestream perturbations (ZD25). Therefore, a predictive framework is required to quantify the downstream amplitude of non-modal perturbations under specific, physically relevant freestream forcing.} Although direct numerical simulations (DNS) provide  a reliable tool for quantitatively linking the boundary-layer perturbations to the freestream forcing \citep{Zhong2006boundary,wan2018response,guo2025transition}, the substantial computational cost precludes systematic parametric studies.

{\color{cyan}For boundary layers with a sharp leading edge, the boundary-region equations (BRE) provide an efficient framework for characterising the formation and evolution of non-modal perturbations. In the incompressible regime, \cite{Leib1999} employed the linear BRE to describe the formation of non-modal streaks induced by infinitesimal freestream, low-frequency vortical forcing. This approach was subsequently extended to compressible flows \citep{Ricco2007response} and nonlinear regimes \citep{ricco2011evolution,Marensi2017}. More recently, \cite{xu2024excitation} utilized the nonlinear BRE to investigate the formation and evolution of compressible G$\ddot{o}$rtler vortices. However, the BRE approach is inadequate for addressing boundary layers with blunt leading edges, where phenomena such as rapid length scale adjustment and the impact of bow shocks become significant.}

To address these limitations, \cite{zhao2025excitation} developed a shock-fitting harmonic  {linearised} Navier–Stokes (SF-HLNS) method, which directly links excited non-modal perturbations to {the} freestream acoustic, entropy, and vortical forcing. {This} approach is efficient because it uses a harmonic {linearised} system, thereby avoiding time integration, and is accurate due to the use of shock fitting to capture shock–perturbation interactions. Using the SF-HLNS results as initial perturbations, \cite{song2025} introduced a prediction framework for bypass transition: the nonlinear evolution of non-modal perturbations is computed {using} NPSE, and {the} secondary instability growth is estimated via the bi-global analysis based on the resulting nonlinear streaky base flow.

Although the receptivity efficiency of non-modal perturbations to various freestream disturbances has been systematically studied over a wide parameter range using the SF-HLNS approach (ZD25), the underlying physical mechanism governing the receptivity process remains unclear. This includes identifying the dominant factors that determine receptivity efficiency and establishing scaling relations among key parameters. Given that asymptotic analysis provides a rational framework for addressing such questions, this paper performs an asymptotic study to elucidate the mechanism of non-modal receptivity. Based on the resulting dominant factors, a reduced model is also developed to enable rapid estimation of receptivity efficiency.

The rest of this paper is structured as follows. In Section \ref{sec:physical_model}, we introduce the physical model and present the SF‑HLNS calculations for a representative non‑modal receptivity case. Section \ref{sec:asymptotic_theory} provides the asymptotic theory describing the perturbation response to freestream forcing around the centerline, with corresponding results shown in Section \ref{sec:results}. Based on the centerline perturbation obtained from the asymptotic theory, Section \ref{sec:LPSE} outlines a parabolic approach for modeling the linear evolution of non‑modal perturbations in downstream boundary layers. Concluding remarks are given in Section \ref{sec:conclusion}.

\section{Physical model and receptivity calculations}
\label{sec:physical_model}
\subsection{Physical model}
\begin{figure}
\begin{center}
  \includegraphics[width=0.95\textwidth]{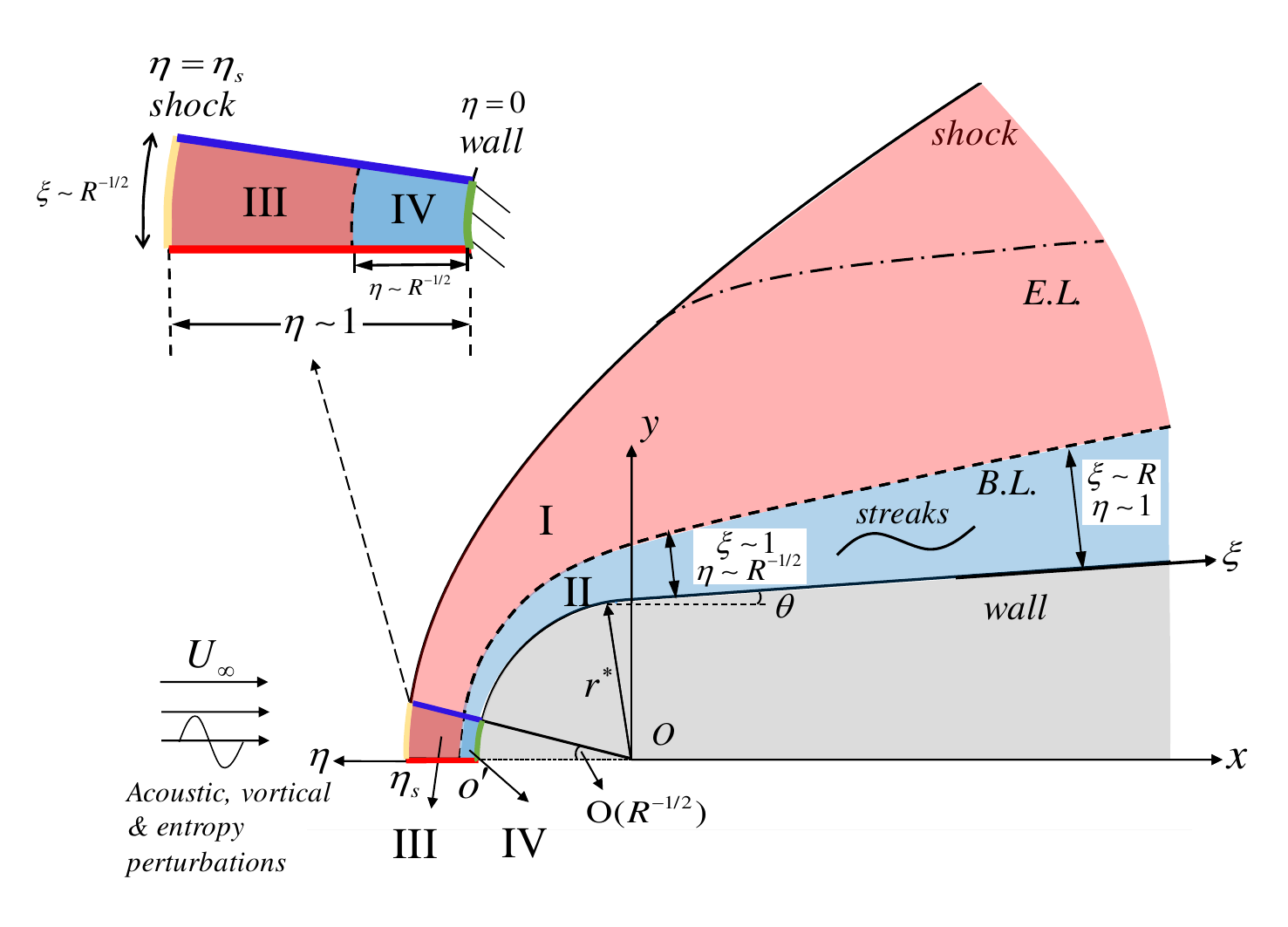}
  \caption{Schematic of the physical problem, where  E.L. and B.L. represent the entropy layer and boundary layer, respectively. Regions I and II correspond to the bulk inviscid flow and the viscous boundary layer, respectively; regions III and IV denote the inviscid and boundary-layer regions around the centreline, respectively. Region II encompasses both the near-nose region $\xi\sim 1$ and the downstream region $\xi\sim R$, for which the boundary-layer thicknesses are $O(R^{-1/2})$ and $O(1)$,  respectively.}\label{fig:sketch}
  \end{center}
\end{figure}
The physical model to be considered is identical to that in ZD25, which is a hypersonic flow passing over a blunt wedge with a semi-angle of  $\theta$ and a nose radius of $r^*$. As sketched in figure \ref{fig:sketch}, a bow shock forms detached from the nose leading edge, and an inviscid entropy layer and a viscous boundary layer emerge.
The reference length is selected to be $r^*$. The dimensionless Cartesian coordinate system $(x,y,z)$ is employed, with its origin locating at the centre of the nose. Using the freestream quantities {\color{red}$(U_\infty^*,\rho_\infty^*,T_\infty^*,\mu_\infty^*)$} as the reference values, we introduce the dimensionless velocity field ${\bf u}\equiv (u,v,w)$, density $\rho$, temperature $T$ and viscosity $\mu$. The pressure is normalised as {\color{red}$p=p^*/(\rho_\infty^* {U_\infty^{*2}})$}. We define two controlling parameters,
 \begin{equation}
 M=U_\infty^*/a_\infty^*,\quad R=\rho_\infty^* {U_\infty^*} r^*/\mu^*_\infty,
 \end{equation}
 where $a_\infty^*$ denotes the freestream sound speed. The Reynolds number in this paper is considered to be sufficiently high, $R\gg 1$. {\color{red}A}lthough the Mach number is numerically high for hypersonic flows, it is treated as  $O(1)$ in the asymptotic analysis.  This ensures that the theoretical framework remains applicable not only to hypersonic conditions but also to supersonic configurations.

 The wall shape is {characterised} by
 \begin{equation}
     y_w=\left\{ \begin{array}{ll}
         \sqrt{1-x_w^2} & x_w<-\sin\theta \\
         \tan\theta x_w+1/\cos\theta & x_w\geq -\sin\theta.
     \end{array}\right.
 \end{equation}
 For better description of the problem, we  additionally introduce a body-fitted coordinate, $(\xi,\eta,z)$, with $\xi$ and $\eta$ {\color{red}aligned} and perpendicular to the body surface, respectively. The origin of this coordinate system is located at the stagnation point; see figure \ref{fig:sketch}. In this paper, the streamwise and wall-normal directions are defined to be along the  $\xi$ and $\eta$ directions, respectively. Consequently,  the velocity vector can be projected to the body-fitted coordinate system, ${\bf u}=(u_\xi,u_\eta,w)$, with $u_\xi$ and $u_\eta$ being the streamwise and wall-normal velocities, respectively.

 The flow is governed by the compressible Navier–Stokes equations. Exploiting the symmetry about the centerline at zero angle of attack, the streamwise computational domain extends from the centerline ($\xi=0$) to a sufficiently downstream location ($\xi=\xi_N$). In the wall-normal direction, the domain is bounded by the wall ($\eta=0$) and the shock location ($\eta=\eta_s(\xi)$), as determined by the shock-fitting technique. {\color{red}The governing equations and boundary conditions are detailed in Section 2.2 of ZD25. Specifically, the system is governed by  (2.8)  supplemented by (2.17), subject to the boundary conditions prescribed in (2.18)–(2.19) and (2.23).}

\subsection{Freestream forcing}
\label{sec:freestream}
The oncoming flow is subject to acoustic, vortical, or entropy perturbations, which can excite boundary-layer perturbations via the receptivity mechanisms. These perturbations may manifest as normal modes (e.g., Mack first and second modes), leading to transition through the natural route, or as non-modal perturbations that take the form of streamwise streaks and trigger transition via the bypass route. This study focuses on the latter, for which low-frequency freestream perturbations are particularly relevant.

 The infinitesimal freestream perturbation in the uniform stream ${\bf u}_\infty=(1,0,0)$ can be expressed as
 \begin{equation}
\frac{\epsilon}{2} {\hat \phi_\infty} \re^{\ri (k_1x+k_2y+k_3z-\omega t)}+c.c.,
\label{eq:freestream_perturbation}
 \end{equation}
 where $\epsilon\ll 1$ denotes its amplitude, ${\bm k}=(k_1,k_2,k_3)$ is the wavenumber vector, $\omega$ is the frequency, and $c.c.$ denotes the complex conjugate. {A} declination angle is introduced, denoted by $\vartheta=\tan^{-1}(k_2/k_3)$.
 The vector $\hat \phi_\infty=(\hat u_\infty,\hat v_\infty,\hat w_\infty,\hat \rho_\infty,\hat \theta_\infty,\hat p_\infty)$ denotes the perturbation state, {\color{red}comprising the velocity components, density, temperature and pressure}. It is normalised such that its energy norm ${\cal E}_\infty$ is unity. Following \cite{mack1969boundary}, we define the energy norm of $\hat \phi_\infty$ as
 \begin{equation}
     {\cal E}_\infty=\frac{|\hat\rho_\infty|^2}{\gamma M^2}+|\hat u_\infty|^2+|\hat v_\infty|^2+|\hat w_\infty|^2+\frac{|\hat\theta_\infty|^2}{\gamma(\gamma-1) M^2}\equiv 1.
 \end{equation}
  In this study, we consider $k_3=O(1)$, which is confirmed to be the most efficient from ZD25.
 From the linearised Euler equations, we can derive the dispersion relation of the freestream perturbations as follows.

 (i) For freestream acoustic perturbations, the dispersion relation is 
 \begin{equation}
     \omega=k_1\pm |\bm k|/M,\label{eq:dispersion_acoustic}
 \end{equation}
 with the eigenfunction being
 \begin{equation}
  \hat \phi_\infty =\frac{1}{\sqrt{2}M}\Big(\pm\frac{M k_1}{|{\bm k}|},\pm\frac{M k_2}{|{\bm k}|},\pm\frac{M k_3}{|{\bm k}|},M^2,(\gamma-1)M^2,1\Big),\label{eq:acoustic_freestream}
  \end{equation}
  where the plus and minus signs denote the fast and slow acoustic wave, respectively.

 (ii) For freestream vortical perturbations, the  dispersion relation is
 \begin{equation}
    \omega=k_1,\label{eq:dispersion_vortical}
 \end{equation}
 with  the eigenfunction being
  \begin{equation}
  \hat \phi_\infty =\Big(\hat u_\infty,\hat v_\infty,\hat w_\infty,0,0,0\Big),
  \end{equation}
  where  $
(\hat u_\infty,\hat v_\infty,\hat w_\infty)$ can be found in (2.31) of ZD25. Here, the vertical vorticity $\hat \Omega_2\equiv k_3\hat u_\infty-k_1\hat w_\infty$ is a free parameter to determine the eigenfunction.

 (iii) The dispersion relation of the freestream entropy perturbation is also (\ref{eq:dispersion_vortical}), and the eigenfunction is
  \begin{equation}
  \hat \phi_\infty =\sqrt{\gamma-1}M\Big(0,0,0,1,-1,0\Big).
  \end{equation}

\subsection{{\color{red}Demonstration of}  receptivity calculations by SF-HLNS}
\label{sec:HLNS_calculations}
The flow field {$\phi=(\rho,u,v,w,T)$} subject to the freestream forcing illustrated in $\S$ \ref{sec:freestream} can be decomposed into a base flow $\bar\phi$ and an infinitesimal perturbation $\tilde\phi$,
\begin{equation}
\label{eq:decompositioni}\phi(\xi,\eta,z,t)=\bar\phi(\xi,\eta)+\epsilon\tilde \phi(\xi,\eta)\re^{\ri(k_3 z-\omega t)}+c.c..
\end{equation}
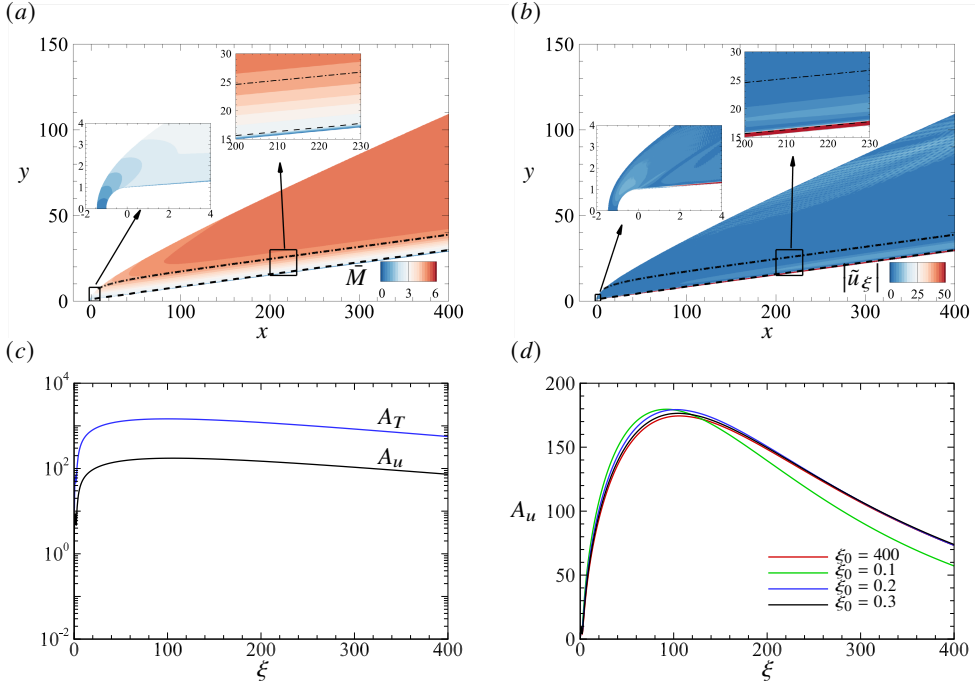
\begin{figure}
  \begin{center}
  \includegraphics[width = 0.49\textwidth]{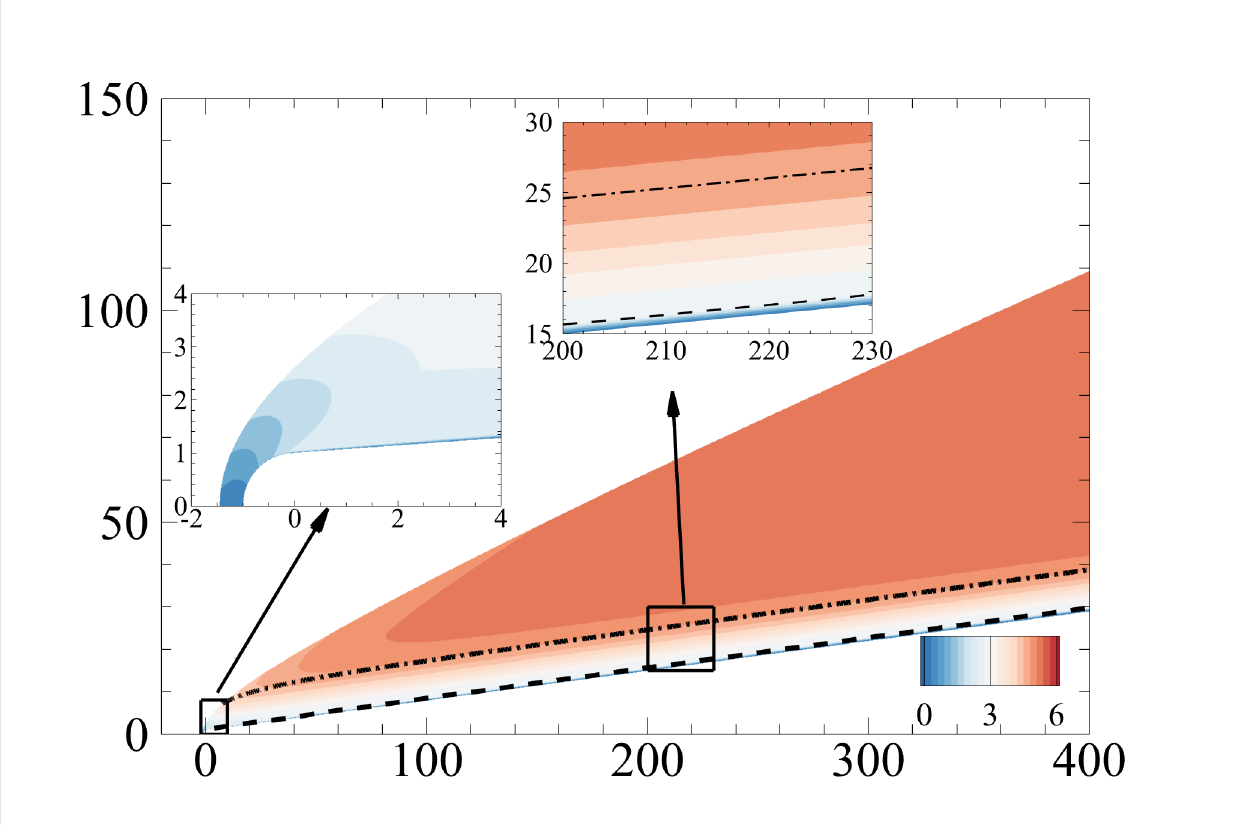}
  \includegraphics[width = 0.49\textwidth]{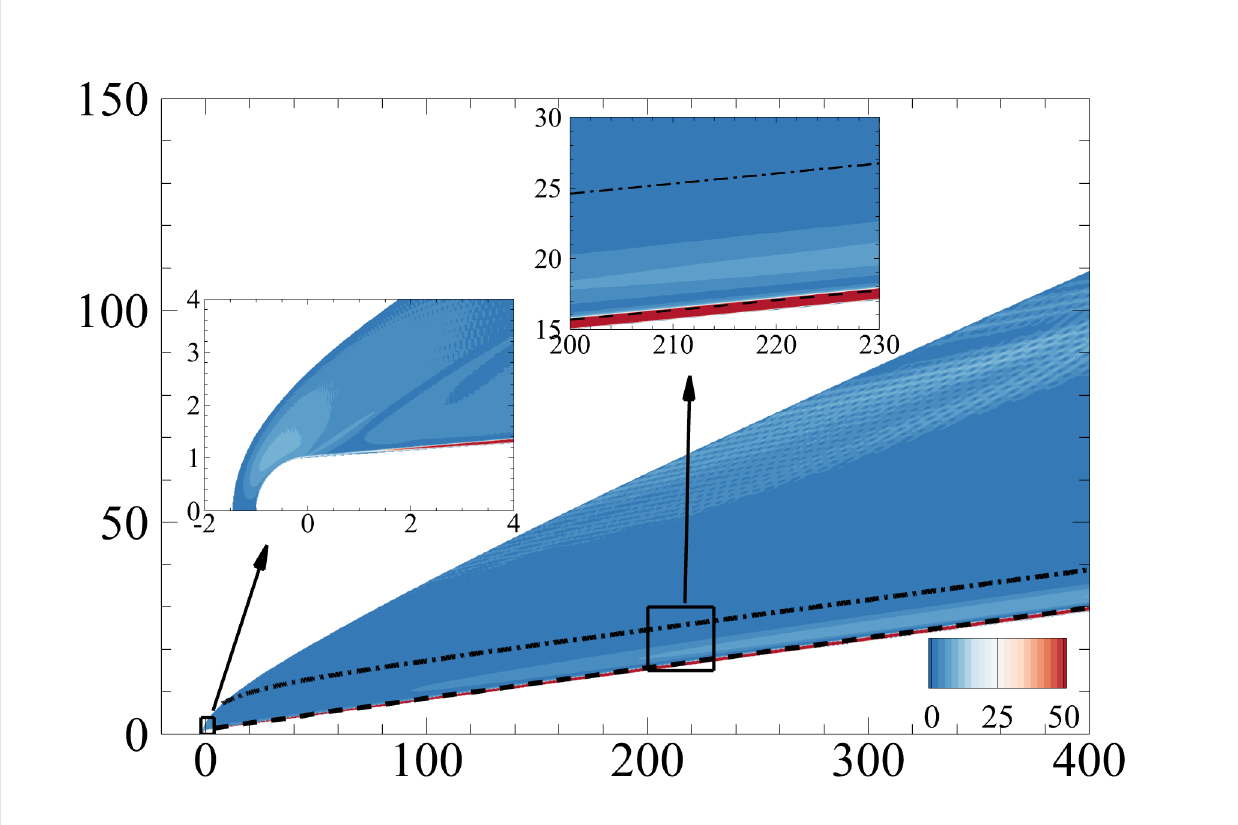}
  \put(-380,120) {$(a)$}
  \put(-190,120) {$(b)$}
  \put(-65,20) {{$\left | \tilde{u } _{\xi}\right | $}}
  \put(-252,20) {$\bar M$}
  \put(-95,0) {$x$}
  \put(-285,0) {$x$}
  \put(-185,60) {$y$}
  \put(-375,60) {$y$}
  \\
  \includegraphics[width = 0.49\textwidth]{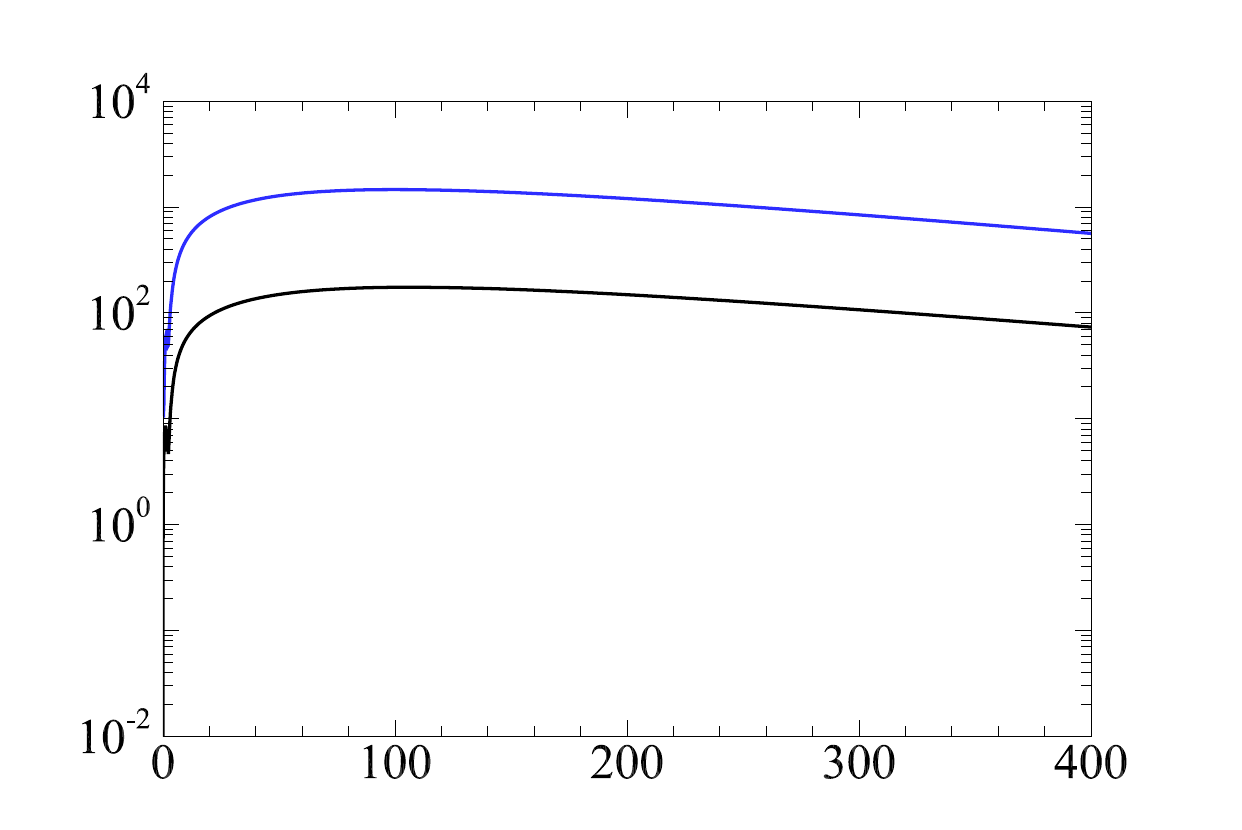}
  \includegraphics[width = 0.49\textwidth]{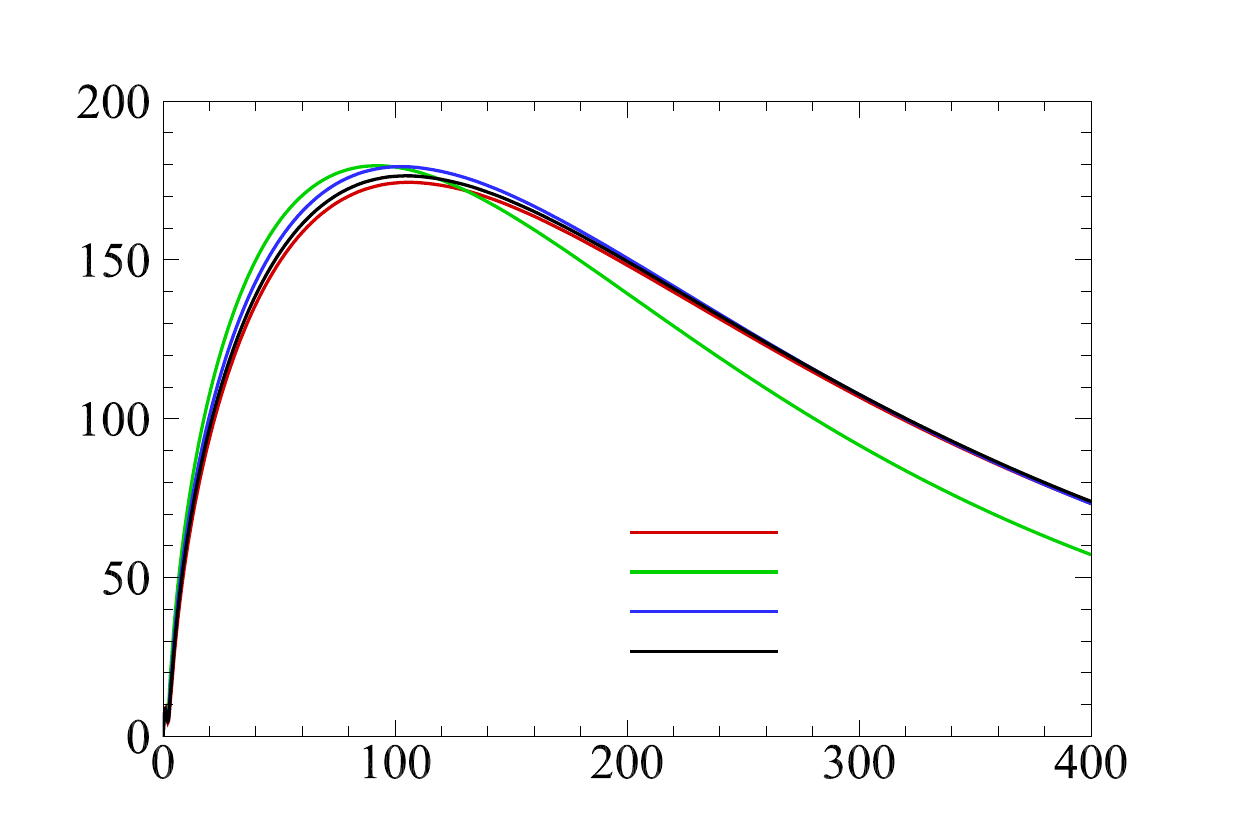}
  \put(-380,120) {$(c)$}
  \put(-190,120) {$(d)$}
  \put(-240,95) {{$A_T$}}
  \put(-240,80) {{$A_u$}}
  \put(-95,0) {$\xi$}
  \put(-285,0) {$\xi$}
  \put(-190,60) {$A_u$}
  \put(-68,44)  {\fontsize{6pt}{6pt}\selectfont$\xi_0=400 $}
  \put(-68,38)  {\fontsize{6pt}{6pt}\selectfont$\xi_0=0.1 $}
  \put(-68,32)  {\fontsize{6pt}{6pt}\selectfont$\xi_0=0.2 $}
  \put(-68,26)  {\fontsize{6pt}{6pt}\selectfont$\xi_0=0.3 $}
  \caption{Contours of the local Mach number $\bar M$ (a) and the perturbation tangetial velocity  ({$\left | \tilde{u}_\xi \right | $}) (b) for $M=5.96$, $T_\infty^*=87$K and {\color{red}$R=167000$}, where the edges of the boundary layer and entropy layer are marked by dashed and dash-dotted lines, respectively. In (b), the freestream forcing is the entropy perturbation with $\omega=0$, {$\vartheta= 0^{\circ}$} and $k_3=8$. (c) Streamwise evolution of the perturbation amplitude ($ A_u $ and  $ A_T$) under entropy forcing. (d) Evolution of $A_u$ with entropy forcing in the region $\xi\in[0,\xi_0]$, where the baseline case denotes  $\xi_0=400$.}
  \label{fig:baseflow}
  \end{center}
\end{figure}
The base flow $\bar\phi$ is computed  using an in-house shock-fitting DNS code, while the perturbation field is obtained via the SF-HLNS  approach; detailed numerical procedures are described in ZD25. 
{\color{red}For a given wall shape, the governing parameter space for  non-modal receptivity includes the Mach number $M$, the Reynolds number $R$, the freestream temperature $T_\infty^*$, the wall temperature $T_w$, and the freestream perturbation characteristics (i.e., the perturbation type, its frequency $\omega$, wavenumber $\textbf k$ and declination angle $\vartheta$). A comprehensive parametric study of this space is presented in ZD25. Here, we aim to demonstrate the receptivity calculation under a fixed parameter set: specifically, $M=5.96$, {$T_\infty^*=87$ K} and $R=167000$ and an adiabatic wall.} {\color{cyan}Notably, the adiabatic-wall condition represents the simplest scenario for illustrating the fundamental receptivity mechanism. Additional cases covering a broader range of   Reynolds numbers and wall temperatures are examined in Part III of this work series.}
Figure \ref{fig:baseflow}-(a) shows the contours of the local Mach number of the base flow, $\bar M=\sqrt{\bar u^2+\bar v^2}M/\sqrt{\bar T}$.  A strong gradient of  $\bar M$  is observed in the entropy layer. In the near-wall boundary-layer region, the local Mach number drops below unity, indicating the presence of a subsonic zone.

It was shown in {\color{red}figure 20 of} ZD25 that for hypersonic blunt boundary layers, freestream acoustic and entropy perturbations produce receptivity efficiencies of comparable magnitude, both significantly stronger than that induced by freestream vortical forcing. Furthermore, the receptivity efficiency decreases with increasing frequency.
Accordingly,  we select an entropy perturbation with $\omega=0$, {$\vartheta= 0^{\circ}$} and $k_3=8$ as representative  demonstration. The resulting perturbation tangential {\color{magenta}velocity} $|\tilde u_\xi|$ is presented in figure \ref{fig:baseflow}-(b), displaying a significant peak in the near-wall boundary-layer region. Although not shown, the excited non-modal perturbation in the downstream region shows longitudinal streaky structures, with the magnitude of {wall-normal} and spanwise velocity perturbations much smaller than that of the longitudinal velocity perturbation; more detailed numerical results with various wall temperatures and nose radii can be found in \cite{Sun2026}.

At each streamwise position $\xi$,  the amplitudes of perturbation streamwise velocity and temperature are defined as {$A_u(\xi):=\max_\eta|\tilde u_\xi(\xi,\eta)|$ and $A_T(\xi)=\max_\eta |\tilde T(\xi,\eta)|$}, respectively. {\color{red}Their streamwise evolution is displayed in figure \ref{fig:baseflow}-(c). Linear stability analysis confirms the absence of unstable normal modes at this frequency and spanwise wavenumber, indicating that the amplitude amplification observed for  $\xi<80$ corresponds to the transient growth of}  {\color{cyan}the excited non-modal perturbation}.  This non-modal perturbation subsequently saturates at  around $\xi\approx 80$. Notably, under a freestream entropy forcing with a unity energy norm, the downstream amplitudes reach  $O(10^2)$ for velocity and  $O(10^3)$ for temperature.

After testing freestream forcing in several localised regions upstream of the bow shock, we identified the  vicinity of the centerline as the most sensitive region for the non-modal receptivity. Figure \ref{fig:baseflow}-(d) shows the streamwise evolution of $A_u$   for freestream entropy forcing applied in the region $\xi\in[0,\xi_0]$ {\color{red}and $\eta=\eta_s$}. The computational domain spans from $\xi=0$ to $\xi_N=400$, and the baseline case   corresponds to $\xi_0=400$, for which the entropy forcing is imposed across the entire upper boundary.
Cases with $\xi_0=0.1$, 0.2 {\color{magenta}and 0.3} represent forcing confined near the centerline. {\color{magenta}The curves for $\xi_0=0.2$ and 0.3 exhibit} good agreement with the baseline case, {\color{magenta}particularly in the downstream region}, while the curve for $\xi_0=0.1$ {underestimates} the downstream perturbation amplitude with an error of only $20\%$. These observations indicate that the perturbation response in a narrow region around the centreline, e.g., $\xi\in[0,0.2]$, dominates the  formation of downstream non-modal perturbations. This process involves perturbation amplification in both the post-shock inviscid region (region III) and the viscous boundary layer near the stagnation point (region IV). The underlying physical mechanism will be uncovered using asymptotic analysis in the following section.

\section{Asymptotic theory and key receptivity mechanisms around the centreline}
\label{sec:asymptotic_theory}

To elucidate the non‑modal receptivity mechanism under different freestream forcing conditions, a high-$R$ asymptotic analysis is employed to examine perturbation excitation in the centerline region. As illustrated in {\color{red}f}igure~\ref{fig:sketch}, four distinct asymptotic layers are identified: the outer inviscid region (I), the main boundary layer (II), the centerline inviscid region (III), and the stagnation‑region boundary layer (IV). Classical boundary‑layer theory, applicable to regions I and II, remains valid everywhere except in the immediate vicinity of the centerline. Although an entropy layer, a region of high density gradient, forms within region I, it does not alter the governing equations or boundary conditions in that region.

{\color{cyan}
To formulate the complete receptivity system, the solution procedure is decomposed into the following sequential steps:
\begin{itemize}[leftmargin=4em]

\item[\textbf{Step (i) \ }] Solve the Euler equations for the outer inviscid region (region I);

\item[\textbf{Step (ii) }] Solve the centerline inviscid equations (region III), utilizing the centerline pressure gradient $dp/d\xi$ obtained from Step (i);

\item[\textbf{Step (iii) }] Solve the viscous system in region IV, employing the lower-limit behavior of the solutions from Step (ii) as the upper boundary condition;

\item[\textbf{Step (iv) }] Solve the downstream boundary-layer equations (region II), using the results from Step (iii) as initial perturbations and those from Step (i) as the upper boundary condition.
\end{itemize}
These steps are applicable to both the base flow and perturbation calculations. However, certain simplifications arising from the flow physics can be implemented.

In the following two subsections, we will focus on step (i) to (iii), aiming at revealing the physical mechanism for the perturbation evolution from the free stream to the stagnation region. The subsequent formation of non-modal streaks, step (iv), will be detailed in Section \ref{sec:LPSE}.
}

\subsection{Base flow}
For step (i), the base flow in region I is governed by the Euler equations, which are solved using our shock‑fitting DNS code by neglecting viscous terms and allowing a slip boundary condition at the wall. {\color{cyan}Consequently, this calculation is independent of both the Reynolds number and the wall temperature; for a prescribed wall geometry, the solution depends solely on the oncoming Mach number.}
{\color{cyan}The mathematical formulations for steps (ii) and (iii) are presented in sections \ref{sec:baseflow_III} and \ref{sec:baseflow_IV}, respectively.
}

A balance of  convection and viscous terms in the boundary-layer region indicates that the thickness of region IV scales as $\eta\sim R^{-1/2}$. Moreover, the key distinction  between regions II and IV lies in the elliptic nature of region IV, where the streamwise and wall-normal length scales are comparable. It follows that  the streamwise extent of region IV scales as $\xi\sim R^{-1/2}$, which  also characteri{\color{red}s}es the streamwise scale of region III.

\subsubsection{Base flow in region III}
\label{sec:baseflow_III}
In region III, we introduce a local coordinate
\begin{equation}
    X=R^{1/2}\xi=O(1).
\end{equation}
In the centreline region,  the bow shock can be approximated as a normal shock. According to the Rankine–Hugoniot (R-H) relations, the post-shock temperature and pressure are elevated by a factor of  $O(M^2)$, while the velocity and density remain of order unity. Given that the dimensionless freestream velocity, density,  temperature and pressure are   $(1,1,1,(\gamma M^2)^{-1})$,  the post-shock velocity, density, and pressure in region I are of $O(1)$, whereas the temperature becomes  $O(M^2)$. 

{\color{red}Since $\xi\ll 1$ in region III, we apply the Taylor expansion to the mean flow $\bar \phi$, 
\begin{equation}
\bar \phi(\xi,\eta)=\bar\phi(0,\eta)+\partial_\xi\bar\phi(0,\eta)\xi+\partial_{\xi\xi}\bar\phi(0,\eta)\frac{\xi^2}{2}+O(\xi^3)\quad\mbox{for }\xi\ll 1.
\end{equation}
Transforming back to the local coordinate $X$, this expansion becomes
\begin{equation}
\bar \phi(X,\eta)=\bar\phi(0,\eta)+\partial_\xi\bar\phi(0,\eta)R^{-1/2}X+\partial_{\xi\xi}\bar\phi(0,\eta)\frac{X^2}{2R}+O(R^{-3/2})\quad\mbox{for }R\gg 1.
\end{equation}
From the symmetric conditions imposed at $\xi=0$, we deduce that
\begin{equation}
    \bar u_\xi(0,\eta)=\partial_\xi\bar u_\eta(0,\eta)=\partial_\xi\bar\rho(0,\eta)=\partial_\xi\bar T(0,\eta)=\partial_\xi\bar P(0,\eta)=0.
\end{equation}
}
Consequently, the base flow in region III is approximated by
\begin{equation}
(\bar u_\xi,\bar u_\eta,\bar w,\bar \rho,\bar T,\bar P)=\Big(\frac{XU_1(\eta)}{R^{1/2}},V_0(\eta),0,R_0(\eta),M^2T_0(\eta),P_0(\eta)+\frac{X^2P_2(\eta)}{2R}\Big)+\cdots,
\label{eq:base_flow_quantities}
\end{equation}
{\color{red}where $U_1=\partial_\xi\bar u_\xi(0,\eta)$, $V_0=\bar u_\eta(0,\eta)$, $R_0=\bar \rho(0,\eta)$, $T_0=M^{-2}\bar T(0,\eta)$, $P_0=\bar P(0,\eta)$ and $P_2=\partial_{\xi\xi}\bar P(0,\eta)$. Notably, the pressure expansion is retained to second order to account for the small magnitude of the streamwise momentum equation, and the temperature is rescaled by $M^2T_0$ to prevent the appearance of numerically large quantities.}
In the curved coordinate, the Lame coefficients in the three directions are $H_1=1+\kappa(\xi)\eta$ and $H_2=H_3=1$, where $\kappa(\xi)$ denotes the local curvature of the body contour. Because the nose radius is selected to be the reference length, the curvature here is  $\kappa(\xi)=1$ for $\xi\in[0,\pi/2-\theta)$. Then, in the centreline region, we have   $H_1=1+\eta$, and the steady Navier-Stokes (N-S) equations  {\color{red}are reduced to}
\refstepcounter{equation}
$$
R_0U_1+(1+\eta)(R_0V_0'+V_0R_0')+ R_0V_0=0,\quad R_0[U_1^2+(1+\eta)V_0U_1'+ U_1V_0]+P_2=0,
\eqno{(\theequation a,b)}
$$
$$
R_0V_0V_0'+P_0'=0,\quad R_0T_0'-(\gamma-1)P_0'=0,\quad R_0T_0=\gamma P_0,\eqno{(\theequation c,d,e)}\label{eq:EQ_inviscid_0}
$$
where throughout this paper, a prime denotes derivative with respect to its argument.
Eliminating $R_0$ and $P_0$, we can recast the equation system to
\refstepcounter{equation}
$$
U_1+(1+\eta)V_0'+\frac{(1+\eta)V_0T_0'}{(\gamma-1) T_0}+ V_0=0,\quad V_0V_0'+\frac{T_0'}{\gamma-1}=0,\eqno{(\theequation a,b)}
$$
$$
U_1^2+(1+\eta)V_0U_1'+ U_1V_0+\frac{T_0P_2}{\gamma P_0}=0.\eqno{(\theequation c)}\label{eq:EQ_inviscid}
$$
Given that the bow shock at the centreline can be represented approximately as a normal shock, we introduce the R-H relation to determine the upper boundary conditions at $\eta=\eta_s$,
\refstepcounter{equation}
$$
V_0(\eta_s)=-\frac{2+(\gamma-1)M^2}{(\gamma+1)M^2},\quad R_0(\eta_s)=\frac{(\gamma+1)M^2}{2+(\gamma-1)M^2},
\eqno{(\theequation a,b)}$$
$$
P_0(\eta_s)=\frac{1}{\gamma M^2}+\frac{2(M^2-1)}{(\gamma+1)M^2},\quad T_0(\eta_s)=\frac{[2\gamma M^2+1-\gamma][2+(\gamma-1)M^2]}{(\gamma+1)^2M^4}.
\eqno{(\theequation c,d)}
\label{eq:R_H}$$
These conditions are also confirmed by comparing with preliminary DNS results.
 The non-penetration condition is imposed at the lower boundary,
\begin{equation}
V_0(0)=0.\label{eq:base_region_III_BC0}
\end{equation}

 It should be noted that while the governing equations (\ref{eq:EQ_inviscid}), after applying the rescaling introduced in (\ref{eq:base_flow_quantities}), are independent  of $M$, the upper boundary conditions remain $M$-dependent. At  sufficiently high $M$ values, the hypersonic asymptotic regime emerges, in which  the upper boundary conditions reduce to
 \refstepcounter{equation}
$$
V_0(\eta_s)=-\frac{(\gamma-1)}{(\gamma+1)},\quad R_0(\eta_s)=\frac{(\gamma+1)}{(\gamma-1)},
\quad
P_0(\eta_s)=\frac{2}{(\gamma+1)},\quad T_0(\eta_s)=\frac{2\gamma(\gamma-1)}{(\gamma+1)^2}.
\eqno{(\theequation)}
\label{eq:R_H_hypersonic}$$
For the Mach number considered in figure \ref{fig:baseflow}, $M=5.96$, the density $R_0(\eta_s)$ obtained from (\ref{eq:R_H}) is 5.26, whereas the hypersonic asymptotic expression (\ref{eq:R_H_hypersonic}) gives 6. This discrepancy indicates an evident numerical error when applying the hypersonic asymptotic approximation at this Mach number. Therefore, the present analysis focuses on the regime for $M=O(1)$.

{\color{cyan}Notably, this system itself is not self-contained, as it comprises three equations for four unknown quantities $(U_1,V_0,T_0,P_2)$. This implies that the centerline equations cannot be solved in isolation and require additional information from downstream to reflect the body contour, indicative of the elliptic nature of the overall problem. To resolve this, a matching condition with the downstream region-I solution is required. Specifically, the scaled pressure gradient $P_2/P_0$ is extracted from region-I solution to close the system.} {\color{red} Specifically, if $\bar P(\xi,\eta)$ denotes the pressure obtained from the Euler equation in region I, then $P_0$ and $P_2$ are identified as $\bar P(0,\eta)$ and $\partial^2\bar P/\partial \xi^2(0,\eta)$, respectively.} {\color{cyan}This enables the following numerical procedure for the solution of region III to be employed.}

\begin{itemize}
    \item \textit{Set the values of $V_0(\eta_s)$ and $T_0(\eta_s)$ at the shock position from the R-H relation, and give an initial guess of $U_1(\eta_s)$.}

    \item \textit{Calculate the profiles of $(U_1,V_0,T_0)$ by solving (\ref{eq:EQ_inviscid}) using the Runge-Kutta method, from which we can obtain $V_0(0)$.}

    \item \textit{Update the initial guess of $U_1(\eta_s)$ using the Newton iterative method. The iteration terminates until the non-penetration condition $V_0(0)=0$ is satisfied.}
\end{itemize}

  For the same $M$ as in $\S$\ref{sec:HLNS_calculations}, the system is solved numerically, and the resulting profiles of  $(U_1,V_0,T_0,P_0,R_0)$ are shown as the solid lines in figure \ref{fig:base_flow}-(a). In the bulk region, the asymptotic predictions agree well with the DNS results (indicated by circles). However, a noticeable discrepancy appears in the near-wall region for  $U_1$:  the DNS profile exhibits a significant gradient, whereas the asymptotic solution does not. This suggests that viscous effects must be taken into account in the boundary-layer region. Additionally, because an adiabatic wall is assumed, the temperature and density do not show evident gradient in the boundary layer. Thermal boundary layers appear for cold walls, as  studied in \cite{Sun2026}.

{\color{red}Now we analyse the near-wall behaviours of the region-III solution of the base flow. In the limit of $\eta\to 0$, the boundary condition (\ref{eq:base_region_III_BC0}) determines that $V_0\to C_v\eta+O(\eta^2)$, where $C_v=V_0'(0)$. Then, from (\ref{eq:EQ_inviscid_0}c,d,e), we can derive that $P_0'(0)=T_0'(0)=R_0'(0)=0$, which, combined with the continuity equation (\ref{eq:EQ_inviscid_0}a), leads to $U_1\to -C_v$ as $\eta\to 0$.  
Therefore, the asymptotic behaviours of the inviscid solution in the near-wall region read
\begin{equation}
(U_1,V_0,T_0,P_0,R_0)=\Big(-C_v,C_v\eta,T_{00},P_{00},R_{00}\Big)+\cdots\quad\mbox{as }\eta\to 0,\label{eq:Mean_asymp}
\end{equation}
where $(R_{00},T_{00},P_{00})=(R_0(0),T_0(0),P_0(0))$.}

\subsubsection{Base flow in  region IV}
\label{sec:baseflow_IV}
 Since $U_1$ in (\ref{eq:Mean_asymp}) does not vanish at the stagnation point, a viscous boundary layer must be considered. Balancing the convection and viscous terms in the momentum equation shows that the boundary-layer thickness scales as   $O(R^{-1/2})$. 
{\color{red}In region IV, we introduce a local coordinate $Y=R^{1/2}\eta=O(1)$. Following the same argument as for (\ref{eq:base_flow_quantities}), we expand  the region-IV base flow}  as
\begin{equation}
(\bar u_\xi,\bar u_\eta,\bar w,\bar \rho,\bar T,\bar P)=\Big(\frac{X\bar U_1(Y)}{R^{1/2}},\frac{\bar V_0(Y)}{R^{1/2}},0,\bar R_{0}(Y),M^2\bar T_{0}(Y), P_{00}+\frac{X^2  P_{20}}{2R}\Big)+\cdots,
\end{equation}
where the condition of zero pressure gradient within the boundary layer is imposed, and $P_{20}=P_2(0)$.
{\color{red} Substituting into the  N-S equations, we derive
\refstepcounter{equation}$$
\bar R_0(\bar U_1+\bar V_0')+\bar V_0\bar R_0'=0,\quad \bar R_0(\bar U_1^2+\bar V_0\bar U_1')+{P_{20}}-(\bar\mu_0\bar U_1')'=0,\eqno{(\theequation a,b)}
\label{eq:EQ_BL}
$$
$$ \bar R_0\bar V_0\bar T_0'-\frac{(\bar\mu_0\bar T_0')'}{Pr}=0,\quad \bar R_0\bar T_0=\gamma  P_{00},\eqno{(\theequation d,e)}
$$}
where $Pr=0.72$ is the Prandtl number, and  $\bar \mu_{0}=\bar\mu_0(\bar T_0)$ {\color{red}denotes the viscosity law. While our theory is applicable to arbitrary viscosity models, we adopt the Sutherland law for demonstration. Specifically, $\bar\mu_0(\bar T_0)=(1+\tilde C)(M^2\bar T_{0})^{3/2}/(M^2\bar T_{0}+\tilde C)$, where $\tilde C=110.4\mathrm{K} /T_\infty^*$.

From the region-III solution, we obtain the matching conditions,
\begin{equation}
    (\bar V_0,\bar R_0,\bar T_0)\to \Big(C_vY, R_{00},T_{00}\Big)\quad\mbox{as }Y\to \infty.\label{eq:base_matching}
\end{equation}
The wall boundary conditions read
\begin{equation}
    \bar U_1(0)=\bar V_0(0)=0,\quad \left\{\begin{array}{ll}
    \bar T_0'(0)=0&\mbox{adiabatic wall},
    \\
        \bar T_0(0)=T_w & \mbox{isothermal wall},
    \end{array}\right.\label{eq:base_BC0}
\end{equation}
where $T_w$ indicates the specified dimensionless wall temperature.

For an adiabatic wall, considering that $\bar T_0'(0)=0$ and $\bar T_0'(Y\to \infty)\to 0$, we can estimate from (\ref{eq:EQ_BL}d) that $T_0$ is a constant throughout the boundary layer, which, combining with the constant pressure feature and the equation of state, also determines $\bar R_0$ to be a constant. Specifically, the density and temperature profile are expressed as  
\begin{equation}
     \bar R_0(Y)=R_{00},\quad \bar T_0(Y)=T_{00}.
\end{equation}
This implies that the thermal boundary layer in the centreline region is absent, and the hydrodynamic boundary layer reduces to an incompressible-like configuration.
}
Combining  (\ref{eq:EQ_BL}a) and (\ref{eq:EQ_BL}b), we derive a third-order ordinary differential equation,
\begin{equation}
\bar V_0'''+C_{00} \Big((\bar V_0')^2-\bar V_0\bar V_0''\Big)=-\frac{P_{20}}{\mu_{00}},
\label{eq:EQ_BL_v}
\end{equation}
where $C_{00}=R_{00}/\mu_{00}$, with $\mu_{00}=\bar \mu_0(T_{00})$.
Imposing the wall boundary conditions
\begin{equation}
  \bar V_0(0)=\bar V_0'(0)=0  
\end{equation}
 and the matching conditions 
 \begin{equation}
      \bar V_0'\to C_v\quad \mbox{as }Y\to \infty,
 \end{equation}
 we can solve the equation system (\ref{eq:EQ_BL_v}) using the same numerical approach as for (\ref{eq:EQ_inviscid}) in $\S$\ref{sec:baseflow_III}.

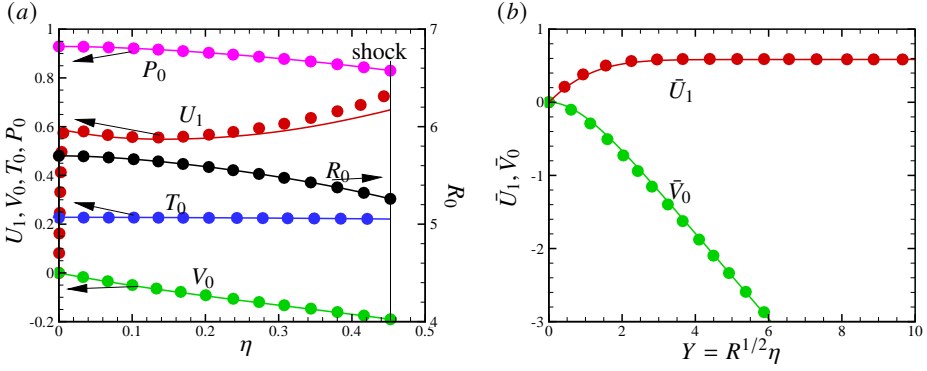
\begin{figure}
\begin{center}
  \includegraphics[width=0.48\textwidth]{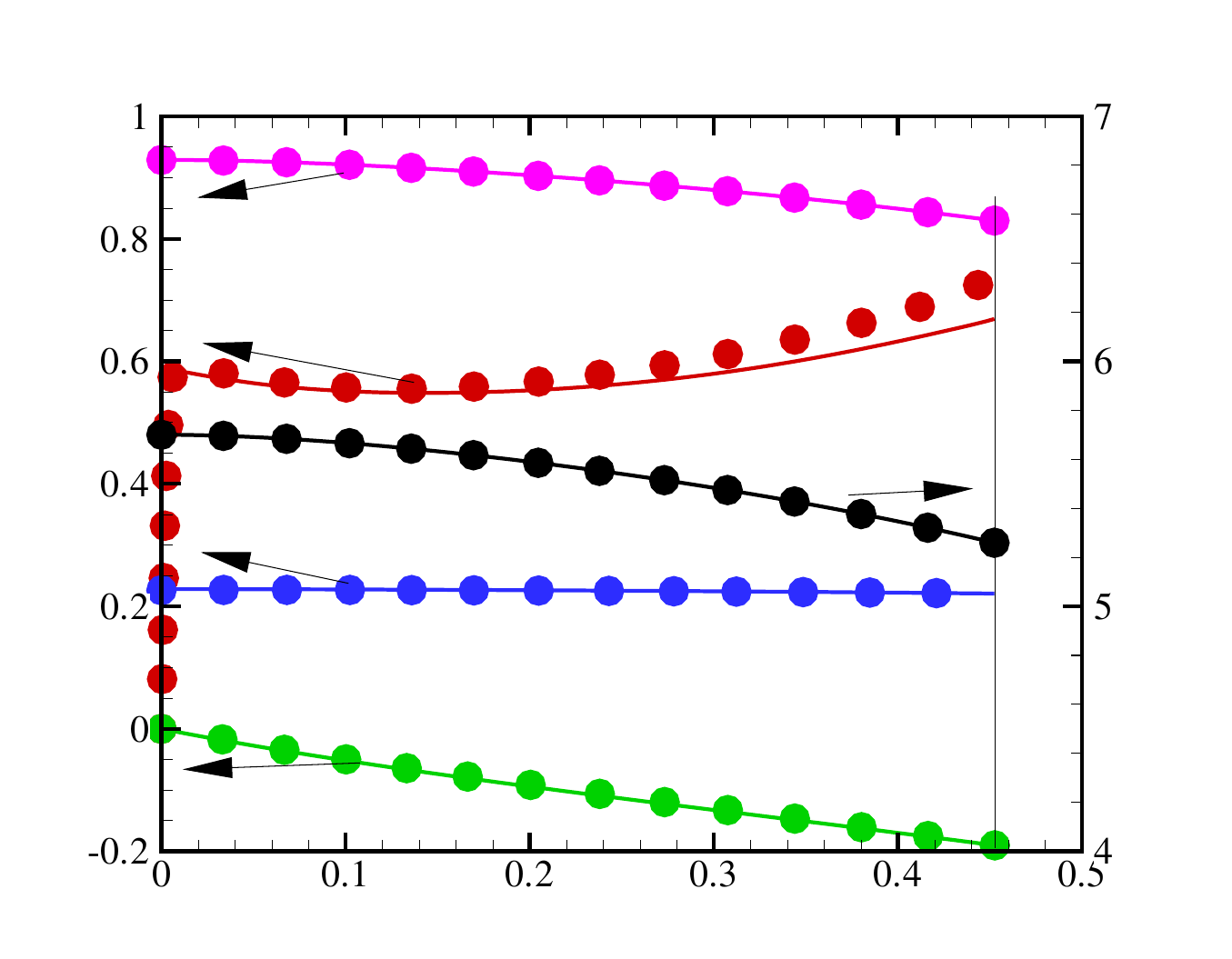}
  \put(-180,130){{$(a)$}}
  \put(-92,5){$\eta$}
  \put(-180,45){\rotatebox{90}{$U_1,V_0,T_0,P_0$}}
  \put(-15,68){\rotatebox{-90}{$R_0$}}
  \put(-115,91){$U_1$}
  \put(-110,30){$V_0$}
  \put(-60,70){$R_0$}
  \put(-50,115){shock}
  \put(-120,58){$T_0$}
  \put(-129,108){$P_0$}
  \includegraphics[width=0.48\textwidth]{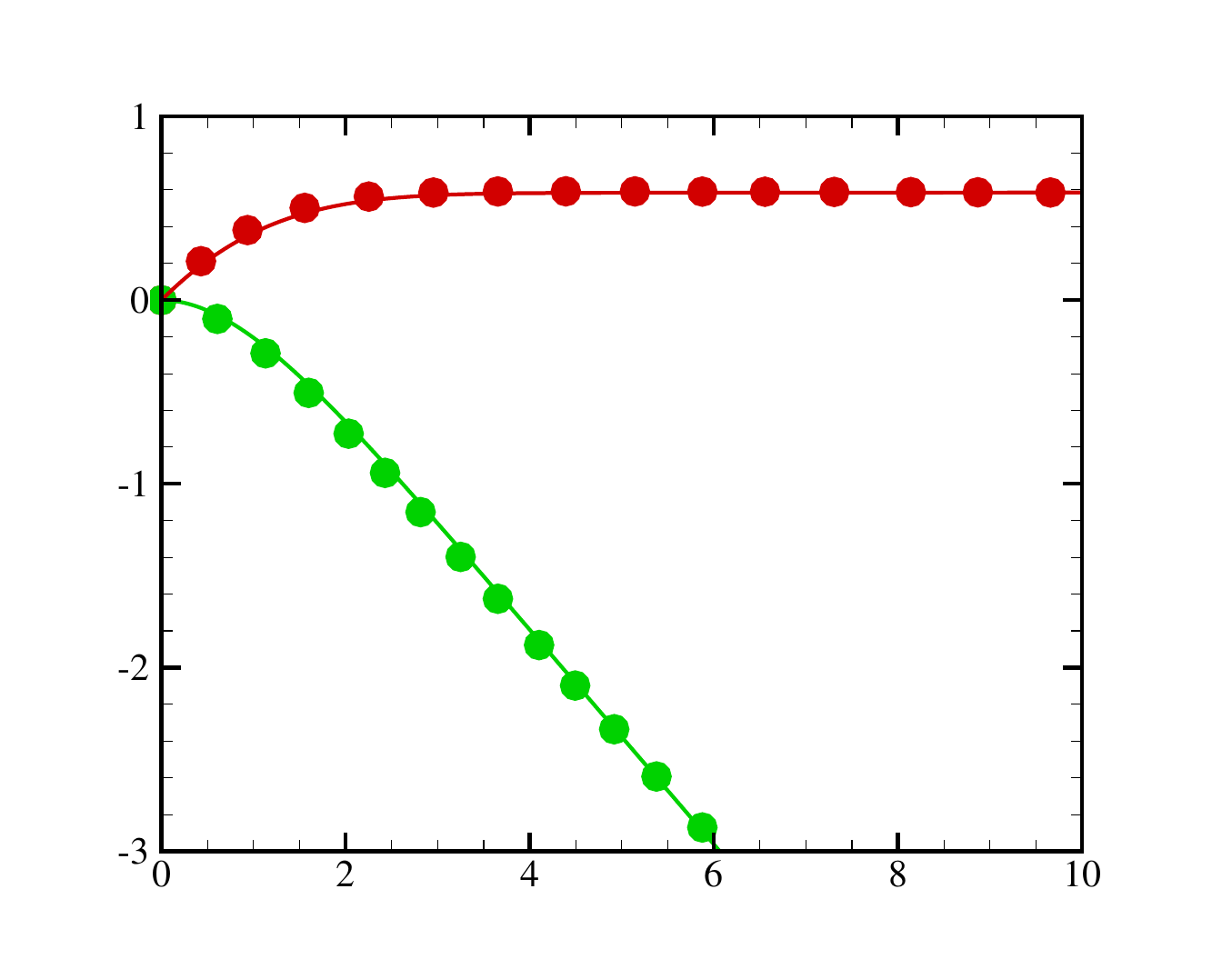}
  \put(-180,130){{$(b)$}}
  \put(-110,2){$Y=R^{1/2}\eta$}
  \put(-180,59){\rotatebox{90}{$\bar U_1,\bar V_0$}}
  \put(-115,100){$\bar U_1$}
  \put(-115,62){$\bar V_0$}
  \caption{Base flow calculations for $M=5.96$. Solid lines: asymptotic prediction; circles: DNS. (a): Bulk region between the wall ($\eta=0$) and the shock $\eta_s=0.452$; (b): boundary-layer region.}\label{fig:base_flow}
  \end{center}
\end{figure}

{\color{red}
For isothermal walls,  the original system (\ref{eq:EQ_BL}) must be solved numerically, subject to the boundary conditions (\ref{eq:base_BC0}) and the matching conditions (\ref{eq:base_matching}). This is achieved using the same numerical procedure for the adiabatic configuration. However, since the primary objective of this paper is to elucidate the key mechanisms of non-modal receptivity, we restrict our verification of the asymptotic analysis to a representative adiabatic case; a comprehensive numerical validation for isothermal walls will be presented in Part III of this series \citep{Sun2026}.
}
Figure \ref{fig:base_flow}-(b) compares the numerical solutions of $\bar U_1$ and $\bar V_0$ with the DNS solutions in the near-wall region, showing excellent agreement.
\subsection{Perturbation field}
Since the boundary-layer response to freestream forcing originates from the centerline region, regions III and IV are of primary relevance. In this subsection, we first {\color{red}analyse} the perturbation field in the inviscid region III under different types of freestream forcing, and then construct the corresponding boundary-layer solution in region IV. The composite solution from these two regions will provide the initial perturbation for the evolution of the boundary layer in region II, as will be discussed in Section \ref{sec:LPSE}.

\subsubsection{Perturbations in region III}
{\color{cyan}ZD25 reported that the wavelengths of non-modal perturbations ($2\pi/k_3$) are comparable to the downstream  boundary-layer thickness (which scales as $O(1)$ for $\xi\sim R$).  Consequently, we adopt $k_3=O(1)$. Furthermore, since the receptivity of  non-modal perturbations intensifies as frequency reduces, and the formation of the downstream streaks require long streamwise length scales, we consider $\omega\ll 1$, (or specifically $\omega=O(R^{-1})$). Under this scaling, the perturbations to leading-order appears to be stationary.} {\color{red}Consistent with the Taylor-expansion principle in (\ref{eq:base_flow_quantities}), the infinitesimal perturbations} in the bulk region are expanded as
\refstepcounter{equation}
$$
\tilde u_\xi=\frac{X\hat u_1(\eta)}{R^{1/2}}\re^{\ri k_3 z}+c.c.+\cdots,
\eqno{(\theequation a)}$$
$$( \tilde u_\eta,\tilde w,\tilde \rho,\tilde T,\tilde p)=(\hat v_0(\eta),\hat w_0(\eta),\hat \rho_0(\eta),M^2 \hat\theta_0(\eta), \hat p_0(\eta)+\frac{\hat p_2(\eta)X^2}{2R})\re^{\ri k_3 z}+c.c.+\cdots.\eqno{(\theequation b)}
$$
The governing equations in this region read
\refstepcounter{equation}
$$
-\frac{V_0R_0'}{R_0}\hat\rho_0+
R_0' v_0+V_0 \hat\rho_0'+R_0\Big(\frac{\hat u_1+\hat v_0}{1+\eta}+\hat v_0'+\ri k_3 \hat w_0\Big)=0,\eqno{(\theequation a)}\label{eq:inviscid_perturbation}
$$
$$
R_0\Big((2U_1+V_0)\hat u_1+[(1+\eta)U_1'+U_1]\hat v_0+(1+\eta)V_0\hat u_1'\Big)-\frac{T_0P_2}{\gamma P_0}\hat\rho_0+{\cal P}\hat p_0=0,\eqno{(\theequation b)}$$
$$
 R_0(V_0\hat v_0)'+V_0V_0'\hat\rho_0+\hat p_0'=0,\quad R_0V_0\hat w_0'+\ri k_3\hat p_0=0,\eqno{(\theequation c,d)}
$$
$$
R_0(V_0\hat\theta_0'+T_0'\hat v_0)+V_0T_0'\hat\rho_0-(\gamma-1)(V_0\hat p_0'+P_0'\hat v_0)=0,\quad \frac{\hat p_0}{P_0}=\frac{\hat\rho_0}{R_0}+\frac{\hat\theta_0}{T_0},\eqno{(\theequation e,f)}\label{eq:EQ_perturbation_III}
$$
where ${\cal P}=\hat p_2/\hat p_0$.
In principle,  the quantity $\hat p_2$ should be prescribed based on the perturbations in region I. {\color{cyan}However, noting that non-modal receptivity is dominated by the perturbation dyamics in the centreline region, the feedback from the downstream region-I perturbation field is negligible. Consequently, the effect of ${\cal P}$ on  receptivity calculations is insignificant. To substantiate  this, we conduct two comparative calculations: one incorporating  ${\cal P}$ from the SF-HLNS calculations, and another  simply setting ${\cal P}=0$.  {\color{red}Under the representative parameter set from figure \ref{fig:baseflow}, the  results of these two calculations are compared  in figure \ref{fig:perturbation}-(a), where they are represented by the dot-dashed and solid lines, respectively.} The excellent agreement between the two results justifies setting  ${\cal P}=0$ as both a convenient and accurate approximation. Crucially, this simplification streamlines the four-step procedure for deriving the perturbation field, as outlined at the beginning of Section \ref{sec:asymptotic_theory}.  Specifically,  step (i) can be removed in the perturbation calculation, and in the subsequent calculations for step (iv), the upper-boundary perturbation can be simply set to zero, given that the dominant contribution arises from the upstream perturbation.}

At the wall, we impose the non-penetration condition, $\hat v_0(0)=0$. At the shock position $\eta=\eta_s$, we need to apply the linearized R-H relations,
\refstepcounter{equation}
$$
V_0\hat\rho_0+R_0\hat v_0=(-\hat \rho_\infty-\hat u_\infty),\eqno{(\theequation a)}
$$
$$
V_0^2 \hat\rho_0+2R_0V_0\hat v_0+\hat p_0=(\hat \rho_\infty+2\hat u_\infty+ \hat p_\infty),\eqno{(\theequation b)}
$$
$$
R_0V_0\hat w_0=\ri k_3(P_0-\frac{1}{\gamma M^2})\hat H,\eqno{(\theequation c)}
$$
$$
\frac{V_0^2(V_0\hat\rho_0+3R_0\hat v_0)}{2}+\frac{\gamma V_0\hat p_0+\gamma P_0\hat v_0}{\gamma-1}
=\Big[-\frac{\hat\rho_\infty+3\hat u_\infty}2-\frac{\hat u_\infty}{(\gamma-1)M^2}-\frac{\gamma \hat p_\infty}{\gamma-1}\Big],\eqno{(\theequation d)}\label{eq:linearized_RH}
$$
where  $\hat H$, appearing in the spanwise momentum equation, denotes the shock movement induced by the freestream forcing.
The left-hand-side terms in (\ref{eq:linearized_RH}a,b,c,d) correspond to the mass, wall-normal momentum, spanwise momentum and energy fluxes  immediately  behind the shock, respectively. {\color{red}These terms are formulated in the body-fitted coordinate system, whereas the right-hand-side terms represent the freestream perturbations, expressed in the Cartesian coordinate system}. Note that the freestream perturbation is shifted by $\re^{-\ri k_1 \eta_s}$ to achieve a brief expression.
The tangential momentum equation is satisfied automatically. Notably, the linearized R-H relations (\ref{eq:linearized_RH}) introduce an additional unknown, the shock movement $\hat H$. To close the system,  the linearized compatibility condition, following (2.39) of ZD25, must be introduced,
\begin{equation}
\Big\{\gamma P_0\Big[\frac{\bar U_1}{\eta_s}+\Big(1-\frac{V_0^2}{T_0}\Big)V_0'\Big]\hat H+\Big[\sqrt{T_0}-(\gamma-1)V_0\Big]\hat p_0+ \frac{(1+\eta_s)\sqrt{T_0}P_0'\hat \rho_0}{R_0}\Big\}_{\eta=\eta_s}=0.\label{eq:linearized_compatibility}
\end{equation}
Thus, the above  system can be solved numerically by a shooting method.

Notably, in the limit of $\eta\to 0$, the base flow $V_0\to C_v\eta$. {\color{red}From (\ref{eq:EQ_perturbation_III}c,e,f), we derive that 
\begin{equation}
    (\hat p_0',\hat \theta_0',\hat \rho_0')=0\quad\mbox{at }\eta=0.\label{eq:perturbation_III_lower_1}
\end{equation}
Subsequently, from (\ref{eq:EQ_perturbation_III}a,d), we can estimate 
}
\begin{equation}
\hat v_0\to -\frac{ k_3^2\hat p_{00}}{R_{00}C_v} \eta\ln \eta, \quad \hat w_0\to  -\frac{\ri k_3 \hat p_{00}}{R_{00}C_v} \ln\eta+B_1,\quad \hat u_1\to \frac{k_3^2\hat p_{00}}{R_{00}C_V}-\ri k_3 B_1\label{eq:asymptote_vw}
\end{equation}
where $\hat p_{00}=\hat p_0(0)$, and $B_1$ is a constant that can be fitted from the bulk-region solution.
For the present freestream entropy forcing, all the perturbation quantities are real, except that $\hat w_0$ is pure imaginary. Therefore, we plot $-\ri \hat w_0$ to show its imaginary part.
Figure \ref{fig:perturbation}-(a) compares the asymptotic predictions (solid lines) with the SF-HLNS results (circles), showing good agreement in the bulk of the layer. The only discrepancy occurs in the near-wall region, {\color{magenta}as highlighted in the zoom-in panel,} where the asymptotic solution for $\hat w_0$ exhibits a divergent trend, consistent with its asymptotic behavior given in (\ref{eq:asymptote_vw}).

 The asymptotic behaviour given in (\ref{eq:asymptote_vw}) further indicates an increase of the streamwise vorticity $\hat \Omega_\xi:=\hat w_0'-\ri k_3\hat  v_0$ like $-\frac{\ri k_3 p_{00}}{R_{00}C_v} \eta^{-1}$ as the wall is approached ($\eta\to 0$). This behaviour is intrinsically driven by a 'slow-down convection' mechanism. {\color{red}Differentiating (\ref{eq:inviscid_perturbation}d) with respect to $\eta$ and  subtracting (\ref{eq:inviscid_perturbation}c) multiplied by $\ri k_3$,} we obtain
 \begin{equation}
 (V_0\hat \Omega_\xi)'=-\frac{R_0'V_0}{R_0}\hat w_0'+\frac{{\ri k_3 V_0V_0'}}{R_0}\hat \rho_0.\label{eq:Omega_inviscid}
 \end{equation}
  As the fluid approaches the stagnation point,  the mean velocity $V_0$ {\color{cyan} tends to vanish, which, under the first-order Taylor's expansion, is approximated by $V_0\sim C_v\eta$ as $\eta\to 0$. Meanwhile, from (\ref{eq:EQ_inviscid_0}c,d,e), we can determine that the inviscid mean-flow solutions at $\eta=0$ behave as $P'_0(0)=T'_0(0)=R'_0(0)=0$.
  Additionally, from (\ref{eq:asymptote_vw}), we can estimate that $\hat w_0'\sim 1/\eta$ as $\eta\to 0$.  Therefore, the right-hand side of (\ref{eq:Omega_inviscid}) decreases like $\eta$ as $\eta\to 0$, indicating that 
  \begin{equation}
       (V_0\hat \Omega_\xi)'\to C_A\eta\quad \mbox{as }\eta\to 0,
  \end{equation}
  where $C_A$ is a constant. The solution is 
  \begin{equation}
     \hat\Omega_\xi\sim \frac{\bar C}{V_0}+\frac{C_A\eta}{2C_v}\quad\mbox{as }\eta\to 0,
  \end{equation}
  where $\bar C$ is a constant. The first term on the right-hand side $\bar C/V_0=\bar C/(C_v\eta)$ diverges as the wall is approached. The physical explanation is that} as the convection speed $V_0$ slows down, the streamwise vorticity perturbation must grow algebraically to maintain the convection balance. This solution remains valid until the boundary-layer region   (in the $O(R^{-1/2})$ vinicity of the stagnation point) is reached, at which the  perturbation streamwise vorticity is amplified by a factor of $O(R^{1/2})$. An indication of this behavior is provided by the circles in figure \ref{fig:perturbation}-(b), which show an algebraic growth of $\Im\{\hat \Omega_\xi\}\equiv -\ri\hat\Omega_\xi$ in the near-wall region, where $\Im$ denotes the imaginary part. This trend is well captured by the inviscid solution for  $\Im\{\hat \Omega_\xi\}$, shown as the solid black line. The agreement remains evident down to $\eta\approx 0.01$, below which viscous effects must be considered.

    This mechanism is rarely observed in classical boundary-layer flows, where the mainstream convects tangentially along the body surface. In such situations, the streamwise vorticity perturbations are typically generated by the entrainment of the vortical perturbations propagating in the free stream \citep{Leib1999,Dong2013continuous}. However, in the nose region of the blunt body, the convective velocity  slows down significantly in the vicinity of the stagnation point, leading to a pronounced enhancement of the  streamwise vorticity perturbation, regardless of the specific type of freestream  perturbations.

 Evidently, strong streamwise vortices are likely to generate longitudinal streaks via the lift-up mechanism, which plays an important role in the downstream evolution of non-modal perturbations \citep{Brandt2014}. Given the  singular behaviour appearing  at the stagnation point $\eta=0$, it becomes necessary to account for the viscous effects in the boundary layer, as studied in the following subsection.

\subsubsection{Perturbations in   region IV}
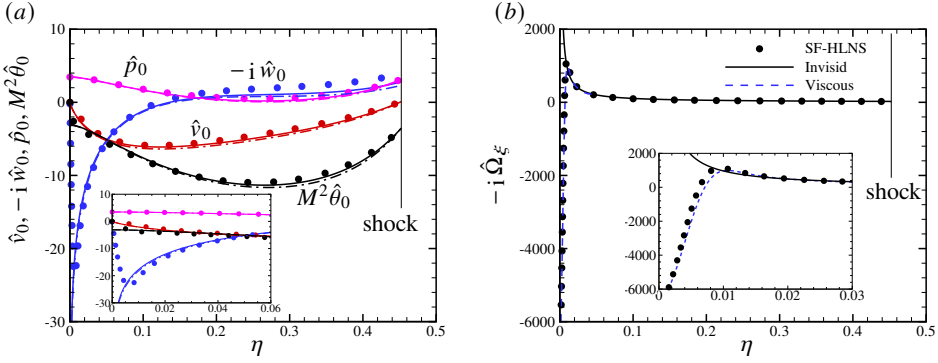
\begin{figure}
\begin{center}
  \includegraphics[width=0.48\textwidth]{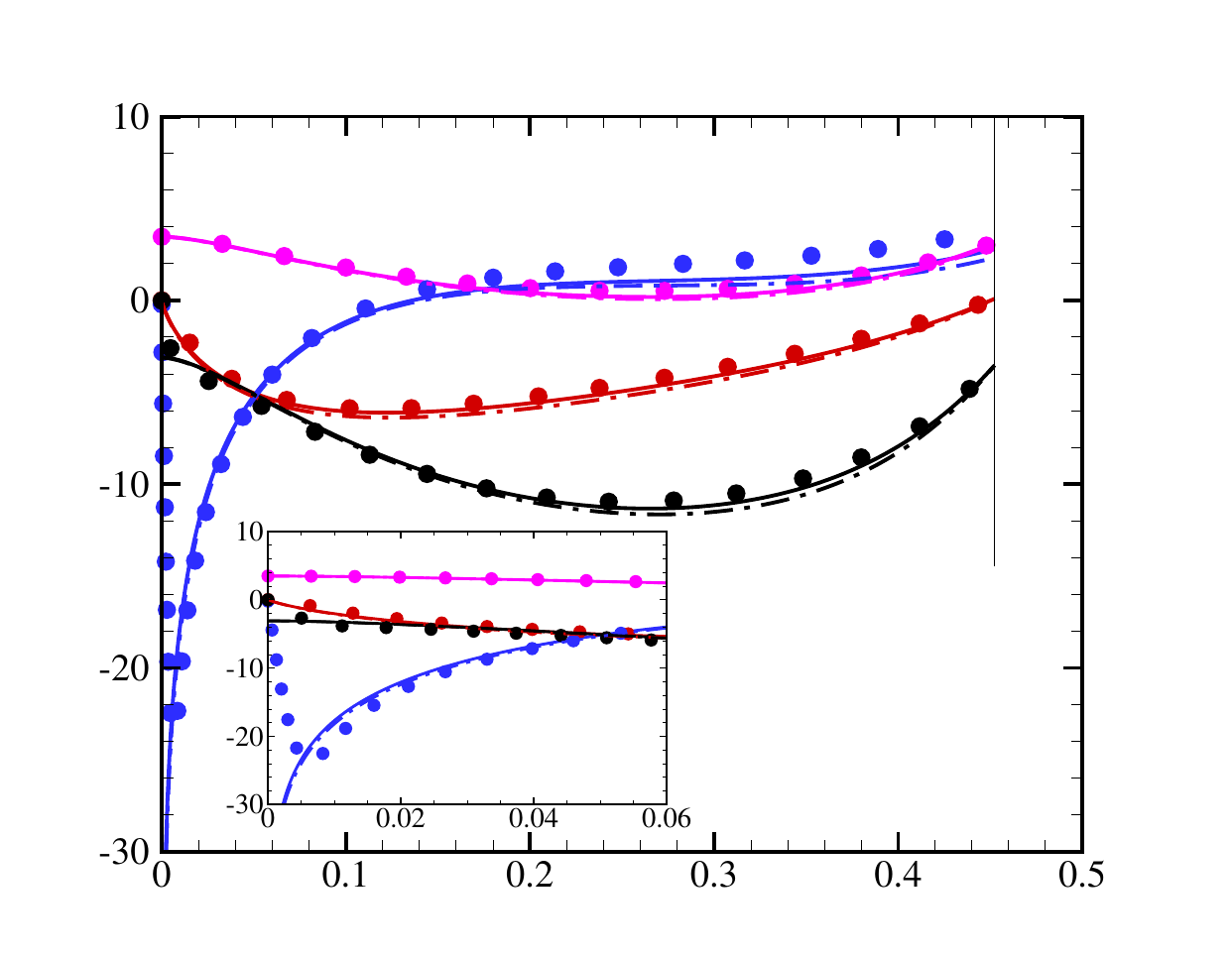}
  \put(-185,130){{$(a)$}}
  \put(-92,5){$\eta$}
  \put(-185,45){\rotatebox{90}{$\hat v_0,-\ri\hat w_0,\hat p_0,M^2\hat \theta_0$}}
  \put(-115,86){$\hat v_0$}
  \put(-50,50){shock}
  \put(-140,110){$\hat p_0$}
  \put(-100,108){$-\ri\hat w_0$}
  \put(-75,60){$M^2\hat\theta_0$}
  \includegraphics[width=0.48\textwidth]{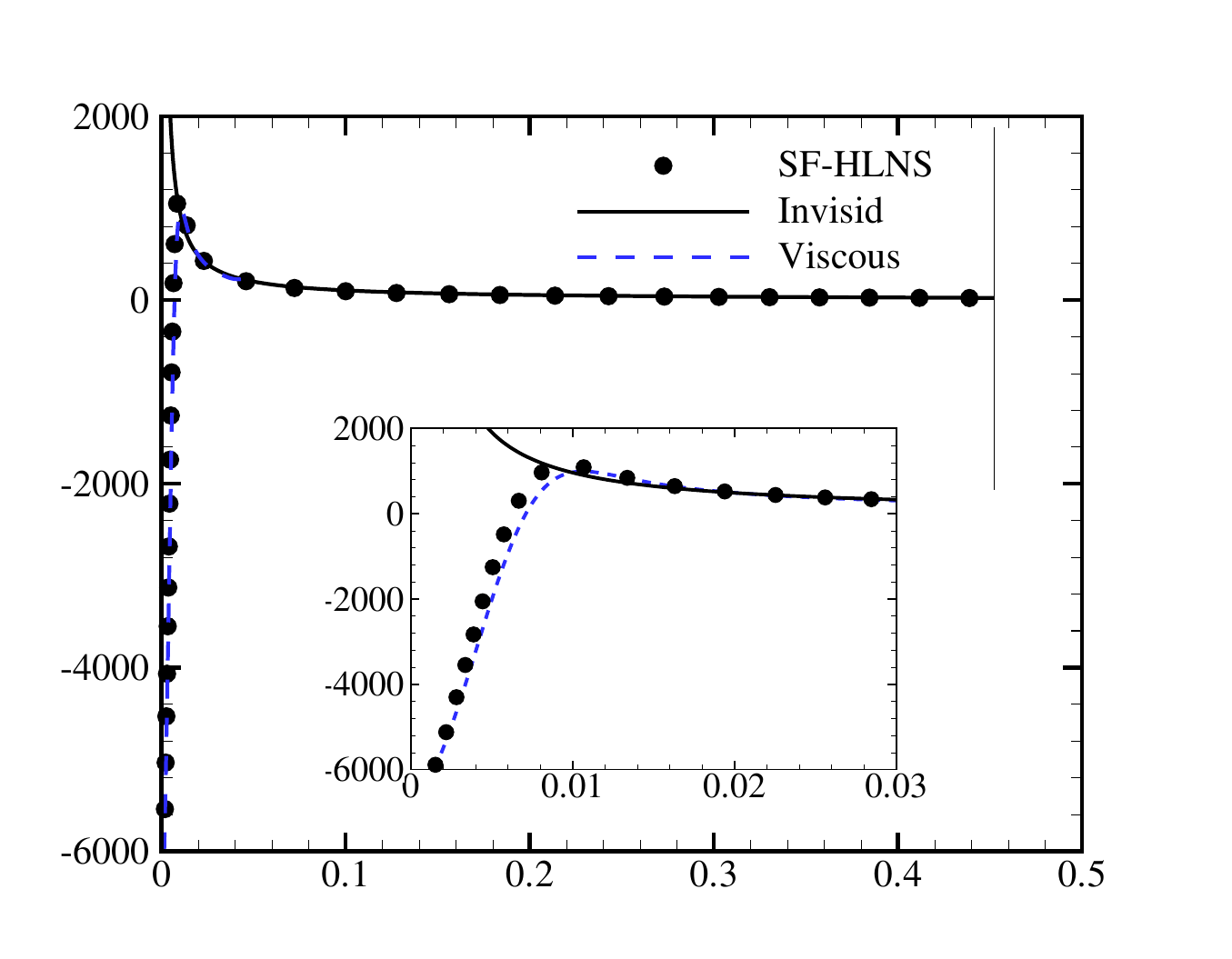}
   \put(-185,130){{$(b)$}}
   \put(-92,5){$\eta$}
   \put(-190,60){\rotatebox{90}{$-\ri\hat\Omega_\xi$}}
   \put(-45,62){shock}
  \caption{Perturbation response to freestream entropy forcing for $M=5.96$,  $\omega=0$, {$\vartheta= 0^{\circ}$} and $k_3=8$. (a): Solutions of ($\hat v_0,\hat w_0,\hat p_0,M^2\hat \theta_0$); (b): solution of $\hat \Omega_\xi$ for $R=167000$. Solid and dashed lines: asymptotic prediction with ${\cal P}=0$; {\color{red}dot-dashed lines in panel (a): asymptotic prediction with ${\cal P}$ given by Region-I solution}; circles: SF-HLNS.  All quantities are real, except that $\hat w_0$ and $\hat \Omega_\xi$ are pure imaginary.}\label{fig:perturbation}
  \end{center}
\end{figure}

To remove the singularity at the stagnation point, the viscous boundary layer  where $Y=O(1)$ is considered. Based on the near-wall behaviours of the region-III solution (\ref{eq:perturbation_III_lower_1}) and (\ref{eq:asymptote_vw}), we expand the perturbations in region IV as 
\begin{equation}
(\hat u_1,\hat v_0,\hat w_0,\hat p_0,\hat \Omega_\xi)=\Big(\bar u_1(Y),R^{-1/2}\bar v_0(Y), \bar w_0(Y),\hat p_{00},R^{1/2}\bar\Omega_\xi(Y)\Big)+\cdots,
\end{equation}
where $\bar\Omega_\xi=\bar w_0'$, and again, the zero pressure-gradient feature in the boundary layer is considered.
 Collecting the leading-order terms, the governing equations are reduced to
 {\color{red}
\refstepcounter{equation}
$$
(\bar U_1+\bar V_0')\bar\rho_0+\bar R_0'\bar v_0+\bar V_0\bar \rho_0'+\bar R_0(\bar u_1+\bar v_0'+\ri k_3\bar w_0)=0,\eqno{(\theequation a)}$$
$$ \bar R_{0}(2\bar U_1\bar u_1+\bar U_1'\bar v_0+\bar V_0\bar u_1')+(\bar U_1^2+\bar U_1'\bar V_0)\bar\rho_0=(\bar \mu_{0}\bar u_1')'+(\mu_T\bar U_1'\bar \theta_0)', \eqno{(\theequation b)}\label{eq:perturbation_BL_iso}
$$
$$\bar R_{0}\bar V_0\bar w_0'+\ri k_3 \hat p_{00}=(\bar\mu_{0}\bar w_0')',\eqno{(\theequation c)}$$
$$
 \bar R_0(\bar V_0\bar\theta_0'+\bar T_0'\bar v_0)+\bar V_0\bar T_0'\bar\rho_0=\frac{1}{Pr}(\bar\mu_0\bar\theta_0'+\mu_T\bar T_0'\bar\theta_0)',\quad \frac{\hat p_{00}}{P_{00}}=\frac{\bar\rho_0}{\bar R_0}+\frac{\bar \theta_0}{\bar T_0},\eqno{(\theequation d,e)}
$$
where {$\mu_T$=d$\bar \mu_0/$d$\bar T_0$}. The wall boundary conditions are
\begin{equation}
    \bar u_1(0)=\bar v_0(0)=\bar w_0(0)=0,\quad\left\{\begin{array}{ll}
         \bar\theta_0'(0)=0&\mbox{adiabatic wall},  \\
        \bar\theta_0(0)=0 &\mbox{isothermal wall}, 
    \end{array}\right.\label{eq:Pert_IV_BC0}
\end{equation}
and the matching conditions are
\begin{equation}
    \Big(\bar u_1',\bar w_0,\bar \theta_0\Big)\to\Big(0,-\frac{-\ri k_3\hat p_{00}}{R_{00}C_v}\ln Y+C_1,\hat \theta_0(0)\Big)\quad\mbox{as }Y\to \infty,\label{eq:Pert_IV_BC1}
\end{equation}
where
\begin{equation}
  C_1=B_1+\frac{\ri k_3 \hat p_{00}}{2R_{00}C_v}\ln R,  \label{eq:C1}
\end{equation}
with $B_1$ being a constant obtained from the inviscid perturbation solution ($B_1=\lim_{\eta\to 0}[\hat w_0+\ri k_3\hat p_{00}\ln\eta/(R_{00}C_v)]$).
 
 If an adiabatic wall is chosen, the mean-flow quantities $\bar R_0$, $\bar T_0$ and $\bar \mu_0$ become constant within region IV, denoted by $\bar R_{00}$, $\bar T_{00}$ and $\bar \mu_{00}$, respectively. From (\ref{eq:perturbation_BL_iso}d,e), we  derive that $\bar \theta_0$ and $\bar \rho_0$ are also constant, taking the values $\hat \theta_0(0)$ and $\hat \rho_0(0)$, respectively. Consequently, the system (\ref{eq:perturbation_BL_iso}) is reduced to
 }
 
\refstepcounter{equation}
$$
\bar u_1+\bar v_0'+\ri k_3\bar w_0=0,\quad R_{00}(2\bar U_1\bar u_1+\bar U_1'\bar v_0+\bar V_0\bar u_1')=\mu_{00}\bar u_1''+{\cal F}_0,\eqno{(\theequation a,b)}\label{eq:perturbation_BL}
$$
$$
R_{00}\bar V_0\bar w_0'+\ri k_3 \hat p_{00}=\mu_{00}\bar w_0'',\eqno{(\theequation c)}
$$
where ${\cal F}_0=-T_{00}P_{20}\hat \rho_0(0)/(\gamma P_{00})$. The boundary and matching conditions are 
\begin{equation}
\bar u_1(0)=\bar v_0(0)=\bar w_0(0)=0,\quad
\bar w_0\to -\frac{\ri k_3 \hat p_{00}}{R_{00}C_v}\ln Y+C_1\quad\mbox{as } Y\to \infty. \label{eq:BC_asym_w}
\end{equation}

From (\ref{eq:perturbation_BL}c), we can derive the solution
\begin{equation}
\bar w_0={\cal W}_p+{\cal C}{\cal W}_c,
\end{equation}
where $\cal C$ is a constant to be determined, and the particular and complementary solutions are expressed as
\begin{equation}
{\cal W}_p=\frac{\ri k_3 \hat p_{00}}{\mu_{00}}\int_0^Y\Big[\re^{C_{00}\int_0^{\bar Y}\bar V_0(\tilde Y)d\tilde Y}\Big(\int_0^{\bar Y}\re^{-C_{00}\int_0^{\breve Y}\bar V_0(\tilde Y)d\tilde Y}d\breve Y\Big)\Big]d\bar Y,
\end{equation}
\begin{equation}
{\cal W}_c=\int_0^Y\re^{C_{00}\int_0^{\bar Y}\bar V_0(\tilde Y)d\tilde Y}d\bar Y,
\end{equation}
where $C_{00}$ was defined in (\ref{eq:EQ_BL_v}).
The upper-limit behaviours of these solutions are
\begin{equation}
{\cal W}_p\to -\frac{\ri k_3 \hat p_{00}}{R_{00}C_v}\ln Y+D_{1p},\quad {\cal W}_c\to D_{1c}\quad\mbox{as }Y\to \infty,
\end{equation}
where $D_{1p}$ and $D_{1c}$ are constants that can be obtained numerically.
Comparing with the matching condition (\ref{eq:BC_asym_w}), we can obtain
\begin{equation}
{\cal C}=(C_1-D_{1p})/D_{1c}.
\label{eq:C}
\end{equation}
Then, the streamwise vorticity can be derived as
\begin{equation}
\bar\Omega_\xi=\bar w_0'=\frac{\ri k_3 \hat p_{00}\re^{C_{00}\int_0^{ Y}\bar V_0(\tilde Y)d\tilde Y}}{\mu_{00}}\int_0^{ Y}\re^{-C_{00}\int_0^{\breve Y}\bar V_0(\tilde Y)d\tilde Y}d\breve Y+{\cal C}\re^{C_{00}\int_0^{ Y}\bar V_0(\tilde Y)d\tilde Y}.\label{eq:Omega_solution_BL}
\end{equation}
The first term on the right-hand side results from the particular solution, which appears to remove the singularity of the inviscid solution and becomes zero at the wall; the second term itself is exponentially decaying in the limit of $Y\to \infty$, it appears as a finite value at the wall. Therefore, the streamwise vorticity perturbation shows a peak at the stagnation point $Y=0$, with the peak value being $\bar \Omega_\xi(0) ={\cal C}$. Although ${\cal C}$ itself is $O(1)$, the magnitude of the streamwise vorticity $\tilde\Omega_\xi:=\tilde w_\eta-\ri k_3\tilde v$ is actually $O(R^{1/2}{\cal C})$, appearing as a significant response at the stagnation point.
It is indicated from (\ref{eq:C}) that  the constant $\cal C$ is determined by matching of the perturbation spanwise velocity between regions III and IV at the $O(1)$ order.  The blue dashed line in figure \ref{fig:perturbation}-(b) represents the boundary-layer solution given by equation (\ref{eq:Omega_solution_BL}), which aligns well with the SF-HLNS results (circles). Therefore, the asymptotic predictions for both the inviscid solution in region III and the viscous solution in region IV are considered reliable.

Eliminating $\bar u_1$ from (\ref{eq:perturbation_BL}a,b), we derive
\begin{equation}
\begin{array}{l}
\vspace{.2cm}
 \bar v_0'''-C_{00}\bar V_0\bar v_0''+2C_{00}\bar V_0'\bar v_0'-C_{00}\bar V_0''\bar v_0  \\ 
    ={\cal F}\displaystyle:=\frac{{\cal F}_0}{\mu_{00}}-\ri  k_3\bar w_0''+\ri k_3 C_{00}(\bar V_0\bar w_0'-2V_0'\bar w_0),
\end{array}
\end{equation}
which is subject to the no-slip boundary conditions {\color{magenta}$\bar v_0(0)=\bar v_0'(0)=0$} and the matching condition {\color{magenta}$\bar v_0\to -\frac{ k_3^2\hat  p_{00}}{R_{00}C_v}Y\ln Y$ as $Y\to \infty$}. Perturbation $\bar u_1$ can be obtained numerically from the continuity equation (\ref{eq:perturbation_BL}).

{\color{red}For isothermal walls, the system (\ref{eq:perturbation_BL}) does not admit an analytical solution. However, numerical solutions can be obtained by solving the linear system subject to the boundary conditions (\ref{eq:Pert_IV_BC0}) and the matching conditions (\ref{eq:Pert_IV_BC1}). Detailed numerical results for these isothermal configurations are presented in Part III \citep{Sun2026}. Nevertheless, the argument regarding the magnitude of the streamwise vorticity perturbation $\bar\Omega_{\xi}$ attains a magnitude of $O(R^{1/2})$ is also valid due to the presence of the thin boundary layer. This indicates a generic amplification mechanism for the streamwise vorticity.}
\subsubsection{Composite solution and discussion}
Once the perturbations in the bulk and boundary-layer regions are obtained, we can construct a composite solution as follows:
\begin{equation}
\left\{\begin{array}{ll}
    \vspace{.2cm}    \tilde u_\xi&=\Big[\hat u_1(\eta)+\bar u_1(R^{1/2}\eta)-\hat u_1(0)\Big]\xi, \\
       \vspace{.2cm}   \tilde u_\eta&\displaystyle=\hat v_0(\eta)+R^{-1/2}\bar v_0(R^{1/2}\eta)+\frac{k_3^2\hat p_{0}(0)}{R_{00}C_v}\eta\ln \eta-b_0\eta, \\ \vspace{.2cm}
         \tilde w&\displaystyle=\hat w_0(\eta) +\bar w_0(R^{1/2}\eta)+\frac{\ri k_3\hat p_{0}(0)}{R_{00}C_v}{\ln\eta}-{\color{magenta}C}_1,\\ \vspace{.2cm}
(\tilde \rho,\tilde \theta)&=\Big(\hat \rho_0(\eta),\hat \theta_0(\eta)\Big)+\Big(\bar \rho_0(R^{1/2}\eta),\bar\theta_0(R^{1/2}\eta)\Big)-\Big(\hat \rho_0(0),\hat \theta_0(0)\Big),\\
\tilde p&=\hat p_0(\eta),
       \end{array}
\right.\label{eq:composite}
\end{equation}
where the constants $b_0$  can be obtained by fitting the asymptotic behaviour of $\bar v_0$  in the large-$Y$ limit. {\color{red}Note that this composite solution is valid for both adiabatic and isothermal walls, although for adiabatic walls, $(\bar\rho_0(R^{1/2}\eta),\bar\theta_0(R^{1/2}\eta)=(\hat\rho_0(0),\hat\theta_0(0))$.}

Based on the preceding analysis, the key mechanism governing receptivity of non-modal perturbations in a blunt-nose boundary layer can be summarized as follows. The detached bow shock interacts with freestream disturbances, generating post-shock acoustic, vortical, and entropy perturbations. These disturbances evolve in the non-uniform inviscid flow, resulting in a redistribution of pressure, vorticity, and entropy fluctuations. In contrast to classical boundary layers, the mean velocity near the centerline decreases significantly toward the stagnation point, exhibiting a strong velocity gradient. Due to the slow-down convection  mechanism, the perturbation streamwise vorticity (i.e., the vorticity tangential to the body surface) increases as  $\eta^{-1}$ when $\eta\to 0$. This growth ceases upon entering the boundary layer, which lies within an  $O(R^{-1/2})$ neighborhood of the stagnation point. As a result, the perturbation streamwise vorticity is amplified by a factor of  $O(R^{1/2})$.  The role of the boundary layer is to remove this singularity through viscous effects. Consequently, a finite, but   $O(R^{1/2})$, perturbation streamwise vorticity develops at the wall. This vorticity perturbation acts as the seed for the formation of downstream longitudinal streaks, as will be demonstrated in Section \ref{sec:LPSE}.

{\color{cyan}Note that the low-frequency configuration is adopted throughout this paper. This choice is necessitated by the requirement for long streamwise length scales that is favourable to the formation of  the non-modal streaks in the downstream region  (region II), a phenomenon corroborated by SF-HLNS calculations (ZD25). Crucially, the slow-down convection mechanism identified in this section does not rely on the small-frequency restriction. For the case where $\omega=O(1)$, unsteady terms enter the leading-order equations, and (\ref{eq:Omega_inviscid}) must be modified to
 \begin{equation}
 -\ri\omega \hat \Omega_{\xi}+(V_0\hat \Omega_\xi)'=-\frac{R_0'V_0}{R_0}\hat w_0'+\frac{{\ri k_3 V_0V_0'}}{R_0}\hat \rho_0.\label{eq:Omega_inviscid_unsteady}
 \end{equation}
 Again, as $\eta\to 0$, this equation reduces to
 \begin{equation}
     -\ri\omega \hat \Omega_{\xi}+(C_v\eta \hat \Omega_{\xi})'\sim 0,
 \end{equation}
 which yields the asymptotic behaviour
 \begin{equation}
     \hat\Omega_\xi\sim \eta^{\ri\omega/C_v-1}\quad\mbox{as }\eta\to 0.
 \end{equation}
Since $|\eta^{\ri\omega/C_v}|=1$, the magnitude of $\hat \Omega_\xi$ diverges like $1/\eta$ as the wall is approached. This singular behavior is again attributed to the deceleration of the convection velocity $V_0$.

}

\section{Predictions of centreline perturbation  using the asymptotic model}
\label{sec:results}
{\color{red}This section aims to elucidate the mechanism governing perturbation evolution in the centerline region. We select a representative case with $M=5.96$, $T_\infty^*=87$K and an adiabatic wall, considering various freestream forcing types and control parameters. The asymptotic predictions are validated against SF-HLNS calculations at $R=167000$. Additional isothermal cases covering a range of Reynolds numbers are detailed in Part III of this series \citep{Sun2026}.}

\subsection{{\color{red}Interactions between freestream perturbations and bow shock}}\label{sec:PSI}
{\color{red}
The first step of the non-modal receptivity process involves the interaction between freestream perturbations and the bow shock. This interaction satisfies the linearized R-H relations, which serve as the upper boundary condition for the Region-III governing equations. Given that the centerline region is the most sensitive in the receptivity process, our analysis focuses on this area where the shock is approximately normal.  We first characterize the free-stream forcing perturbations. 

For \textbf{freestream  acoustic forcing}, considering the low-frequency limit ($\omega\ll 1$),  the dispersion relation (\ref{eq:dispersion_acoustic}) is simplified to
\begin{equation}
k_1=\mp \frac{k_3}{{\cos}\vartheta\sqrt{M^2-1}},\quad |{\bf k}|=\frac{k_3 M}{{\cos}\vartheta\sqrt{M^2-1}},
\end{equation}
where the minus and plus signs denote the fast and slow acoustic forcing, respectively.
The perturbation eigenfunctions (\ref{eq:acoustic_freestream}) are then given by
\begin{equation}
\Big(\hat u_\infty,\hat v_\infty,\hat w_\infty,\hat \rho_\infty,\hat \theta_\infty,\hat p_\infty\Big)=\frac{(-1,\pm{\sin}\vartheta\sqrt{M^2-1},\pm{\cos}\vartheta\sqrt{M^2-1},M^2,(\gamma-1)M^2,1)}{\sqrt{2}M},
\end{equation}
where the   plus and minus signs denote the fast and slow acoustic forcing, respectively. Notably, the perturbations $(\hat u_\infty,\hat p_\infty,\hat \theta_\infty,\hat \rho_\infty)$ are independent of the declination angle and are identical for both fast and slow acoustic forcing.
{\color{red} Since (\ref{eq:linearized_RH}) indicates that the post-shock perturbations are governed solely by $\hat \rho_\infty$, $\hat p_\infty$ and $\hat u_\infty$,} the post-shock perturbations inherit these properties, remaining identical for both fast and slow acoustic forcing types and independent of the declination angle. Furthermore, the density perturbation $\hat \rho_\infty$ is $O(M^2)$ greater than $\hat u_\infty$ and $\hat p_\infty$.

For \textbf{freestream vortical forcing}, the dispersion relation (\ref{eq:dispersion_vortical}) dictates that in the low-frequency limit, the streamwise wavenumber $k_1\approx 0$. The freestream perturbation density $\hat \rho_\infty$ and pressure $\hat p_\infty$ are zero, while the freestream perturbation velocity  $\hat u_\infty=\hat \Omega_2/k_3$. This indicates that the forcing is inversely proportional to $k_3$, but directly proportional to $\hat\Omega_2$. It is important to note that  the dispersion relation requires $|\hat\Omega_2|< |k_3|$ because we have assumed $|\hat u_\infty|^2+|\hat v_\infty|^2+|\hat w_\infty|^2=1$ (and so $|\hat u_\infty|<1$).

For \textbf{freestream entropy forcing}, the dispersion relation remains identical to that for freestream vortical forcing, and only the perturbation density and temperature are non-zero, $\hat \rho_\infty=\sqrt{\gamma-1}M=-\hat\theta_\infty$.

Since the inhomogeneous source terms in (\ref{eq:linearized_RH}) depend only on $\hat\rho_\infty$, $\hat u_\infty$ and $\hat p_\infty$, we compare there quantities for various low-frequency forcing types under a unit energy norm:
 \begin{equation}
\Big({\hat \rho_\infty},{\hat u_\infty},{\hat p_\infty}\Big)\sim \left\{
\begin{array}{ll}
\vspace{.2cm} \displaystyle
   \Big(\frac{M}{\sqrt{2}},-\frac{1}{\sqrt{2}M},\frac{1}{\sqrt{2}M}\Big) & \mbox{for acoustic forcing},\\
   \vspace{.2cm} \displaystyle \Big(0,\frac{\hat\Omega_2}{k_3},0\Big) & \mbox{for vortical forcing},
   \\
   \vspace{.2cm} \displaystyle  \Big(\sqrt{\gamma-1}M,0,0\Big) & \mbox{for entropy forcing}.\end{array}
 \right.
\label{eq:estimate_FS}
 \end{equation}
Crucially, these quantities remain independent of the declination angle $\vartheta$, under the low-frequency assumption ($\omega\ll 1$). 

Based on the linearized R-H relation (\ref{eq:linearized_RH}a,b,d), we can derive the perturbations in the immediate post-shock region as follows:
\refstepcounter{equation}
$$
\hat v_0(\eta_s)=\frac{(\gamma-1)(C_3-C_1V_{0s}^2/2)-\gamma V_{0s}(C_2-V_{0s}C_1)}{\gamma P_{0s}+V_{0s}},\eqno{(\theequation a)}\label{eq:post_shock_perturbations}
$$
$$
\hat \rho_0(\eta_s)=\frac{C_1-R_{0s}\hat v_0(\eta_s)}{V_{0s}},\quad \hat p_0(\eta_s)=C_2-V_{0s}C_1+\hat v_0(\eta_s),\eqno{(\theequation b,c)}
$$
where $(R_{0s},V_{0s},P_{0s})=\Big(R_0(\eta_s),V_{0}(\eta_s),P_0(\eta_s)\Big)$, satisfying the R-H relation (\ref{eq:R_H}), and 
\begin{equation}
    C_1=-\hat \rho_\infty-\hat u_\infty,\quad C_2=\hat \rho_\infty+2\hat u_\infty+ \hat p_\infty,\quad C_3=-\frac{\hat\rho_\infty+3\hat u_\infty}2-\frac{\hat u_\infty}{(\gamma-1)M^2}-\frac{\gamma \hat p_\infty}{\gamma-1}.\label{eq:C123}
\end{equation}
For $M=5.96$, the post-shock mean-flow quantities can be evaluated using (\ref{eq:R_H}) as follows:
\begin{equation}
    R_{0s}=5.260,\quad V_{0s}=-0.1901,\quad P_{0s}=0.8300.
\end{equation}
}

 Evidently, the coefficients $C_1$, $C_2$ and $C_3$ defined in (\ref{eq:C123}) are $O(M)$ for both freestream acoustic and entropy forcing, but are merely $O(1)$ for vortical forcing. This explains the weaker response to vortical forcing from SF-HLNS calculations in ZD25.
  Notably, the prefactors of $\hat \rho_\infty$ and $\hat p_\infty$ in (\ref{eq:linearized_RH}) are all $O(1)$, and those of $\hat u_\infty$  are at most $O(1)$. Consequently,  the  density perturbation $\hat \rho_\infty$, which scales as $O(M)$, emerges as the dominant contributor to the acoustic and entropy receptivity. As illustrated in (\ref{eq:estimate_FS}), the prefactors of $\hat \rho_\infty$,  $1/\sqrt{2}$ for acoustic forcing and $\sqrt{\gamma-1}$ for entropy forcing,  are numerically close,  explaining the comparable receptivity efficiencies for free-stream acoustic and entropy forcing reported in ZD25.

  \begin{table}
  \centering
  \color{red}
  \begin{tabular}{ccccccc}\vspace{.2cm}
  Freestream forcing type &$C_1$&$C_2$&$C_3$ & $\hat {\rho}_0(\eta_s)$ & $\hat v_0(\eta_s)$& $\hat{p}_0(\eta_s)$\\
      Fast acoustic& -4.096&4.096&-2.336 & 20.92 & -0.0226 & 3.294\\
      Slow acoustic& -4.096&4.096&-2.336 & 20.92 & -0.0226 & 3.294\\
    Vortical &-$\hat \Omega_2/k_3$ &2$\hat \Omega_2/k_3$&-1.570$\hat \Omega_2/k_3$& 1.298$\hat \Omega_2/k_3$ & -0.1432$\hat \Omega_2/k_3$ & 1.667$\hat \Omega_2/k_3$\\
      Entropy  &-3.769&3.769&-1.885& 22.27& 0.0884 & 3.141 \\ 
  \end{tabular}
  \caption{Post-shock physical quantities derived by (\ref{eq:post_shock_perturbations}) for $M=5.96$  under freestream entropy, acoustic and vortical forcing.}
  \label{tab:post_shock}
\end{table}

  {\color{red}Using (\ref{eq:post_shock_perturbations}), we compare the post-shock perturbations for various freestream forcing types at $M=5.96$, as summarised in Table \ref{tab:post_shock}. These values are corroborated  by SF-HLNS calculations. Notably, the post-shock responses to  fast and slow acoustic forcing are identical. For both acoustic and entropy forcing, the responses are independent of   the spanwise wavenumber $k_3$; the density perturbation $\hat \rho_0(\eta_s)$ is significantly greater than the velocity $\hat u_0(\eta_s)$ and pressure $\hat p_0(\eta_s)$ perturbations. In contrast,   the response to  vortical forcing is proportional to the vertical vorticity $\hat \Omega_2$ and inversely proportional to $k_3$. Furthermore, the magnitude of perturbation density $\hat \rho_0(\eta_s)$ is comparable with $\hat p_0(\eta_s)$, but is one order of magnitude  smaller than that for acoustic and entropy forcing.}
\subsection{{\color{red}Perturbation evolution towards the stagnation point}}

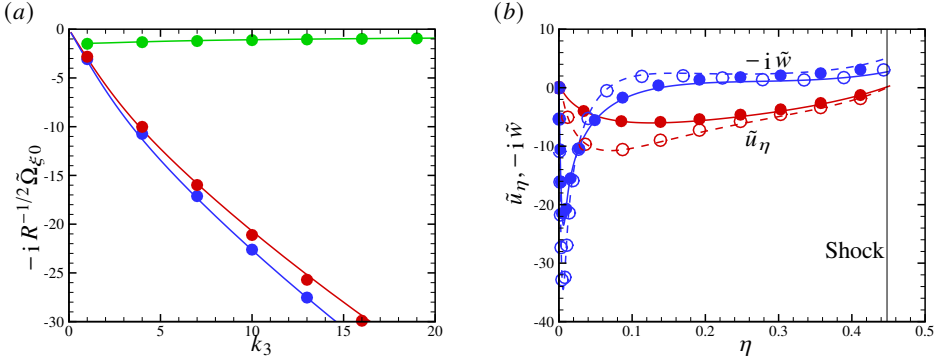
\begin{figure}
\begin{center}
  \includegraphics[width=0.48\textwidth]{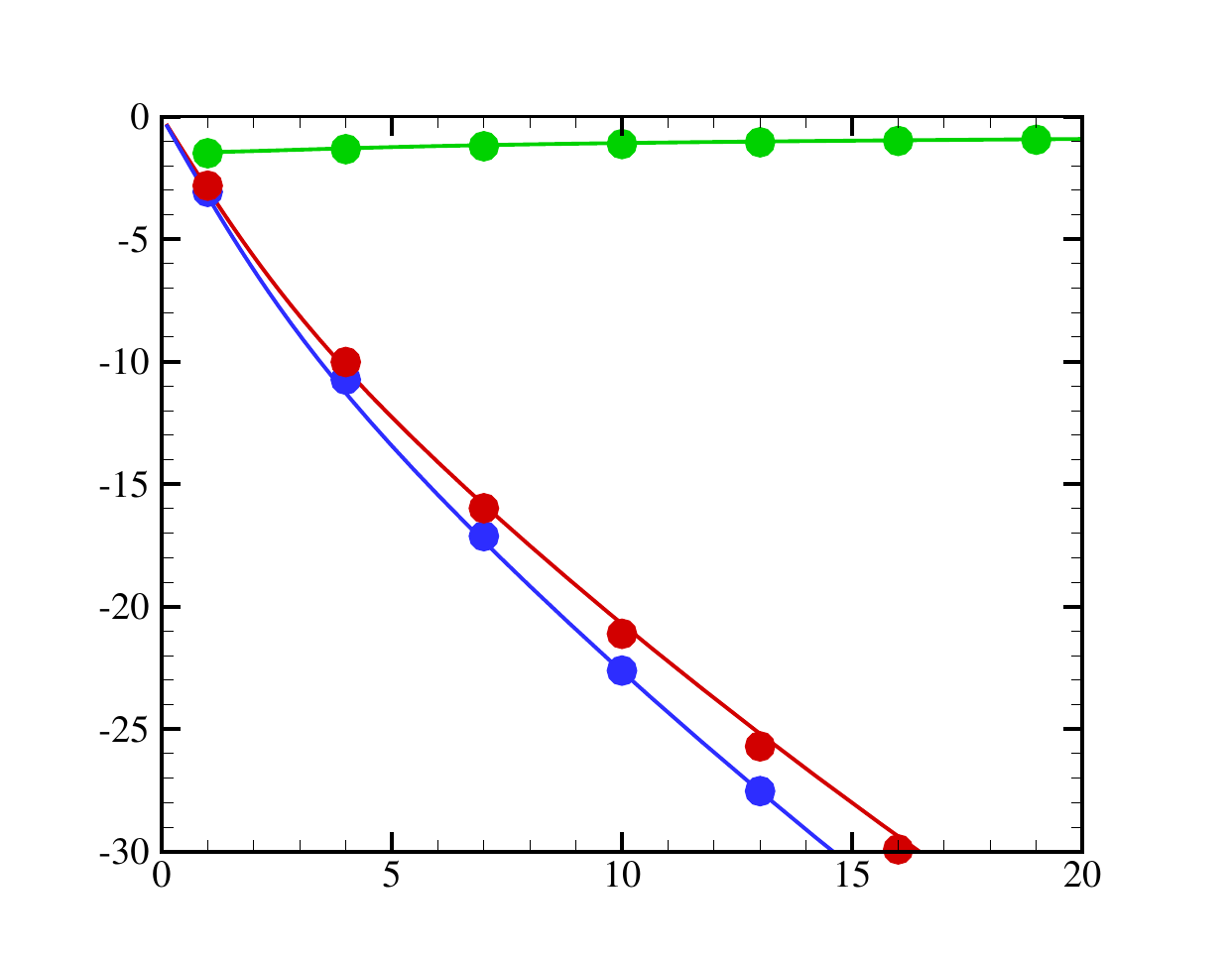}
  \put(-185,130){{$(a)$}}
  \put(-92,5){$k_3$}
  \put(-180,35){\rotatebox{90}{$-\ri R^{-1/2}{\tilde\Omega_{\xi 0}}$}}
  \includegraphics[width=0.48\textwidth]{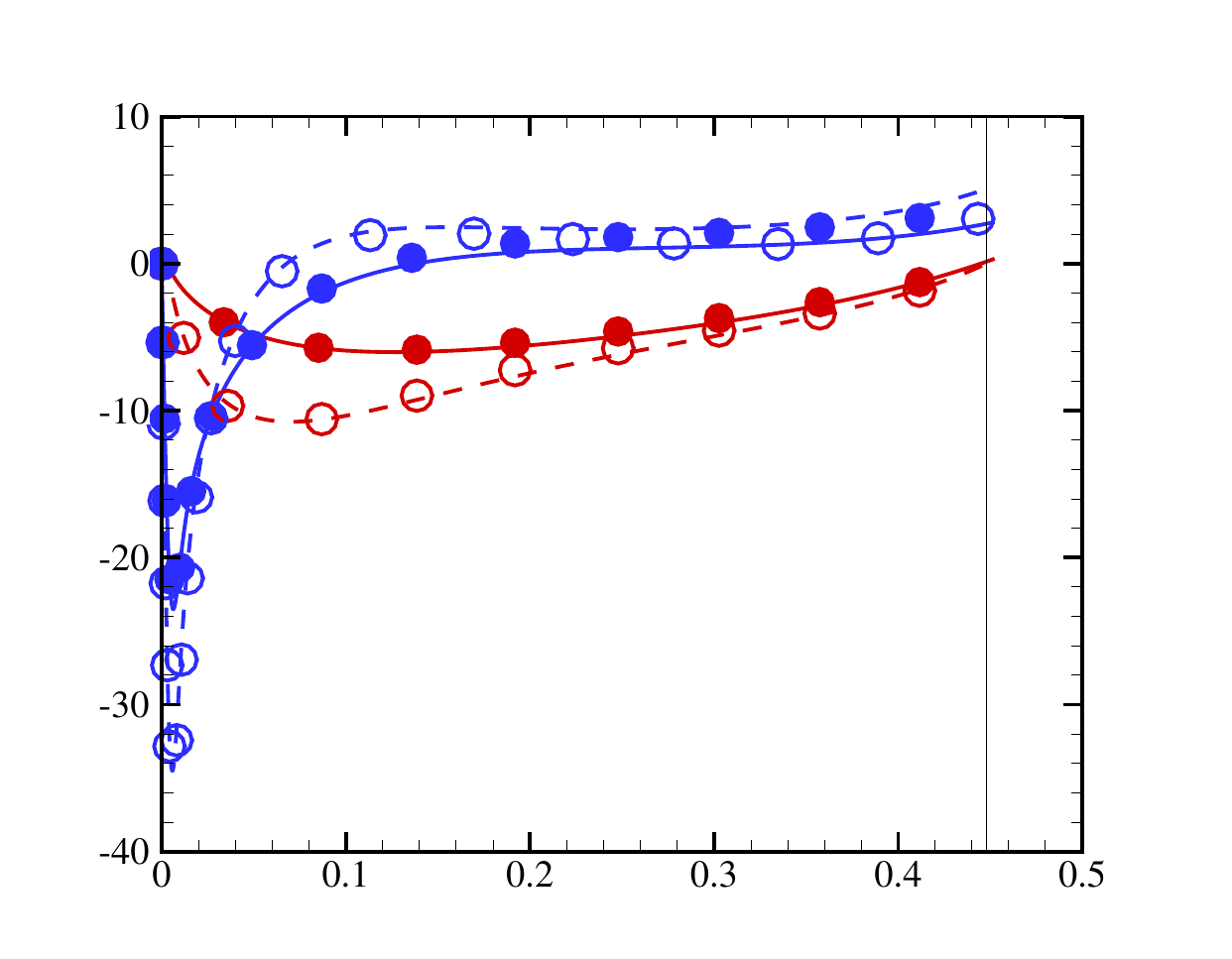}
  \put(-185,130){{$(b)$}}
  \put(-92,5){$\eta$}
  \put(-180,60){\rotatebox{90}{$\tilde u_\eta,-\ri \tilde w$}}
  \put(-60,40){Shock}
  \put(-90,82){$\tilde u_\eta$}
  \put(-90,112){$-\ri \tilde w$}
  \caption{(a) Dependence of the rescaled perturbation streamwise vorticity at the stagnation point on $k_3$ for $M=5.96$ and $T_\infty^*=87$ K under freestream acoustic (blue), vortical (green) and entropy (red) forcing. For vortical forcing, we set $\hat \Omega_2=1.0$.  (b) Profiles of $\tilde u_\eta$ (red) and $\tilde w$ (blue) at the centreline for entropy forcing at $R=167000$, where the solid and dashed lines are for $k_3=8$ and 15, respectively. Continuous lines: asymptotic predictions; symbols:  SF-HLNS calculations. }\label{fig:asymptotic_entropy}
  \end{center}
\end{figure}

The most relevant quantity for demonstrating the slow-down convection mechanism is the perturbation streamwise vorticity at the stagnation point, defined as $\tilde \Omega_{\xi 0}:=\tilde \Omega_\xi(0,0)$. Symmetry dictates that  $\tilde \Omega_{\xi 0}$ is purely imaginary, and the asymptotic analysis reveals that  $\tilde \Omega_{\xi 0}\sim R^{1/2}$. Consequently, we plot the dependence of the spanwise wavenumber $k_3$ on $-\ri R^{-1/2}\tilde \Omega_{\xi 0}$ in   figure \ref{fig:asymptotic_entropy}-{(a)}.  Comparing with the positive  $\Im\{\tilde\Omega_\xi\}$ at the bulk inviscid region, shown in figure \ref{fig:perturbation}-(b), the negative stagnation response of $\Im\{\tilde\Omega_\xi\}$ indicates a phase shift of $\pi$ in the boundary-layer region.

Figure  \ref{fig:asymptotic_entropy}(a)  validates our asymptotic analysis by comparing asymptotic predictions (circles) with SF-HLNS calculations (lines) for  entropy (red), fast acoustic (blue) and vortical (green) forcing. Because the response to slow acoustic forcing is exactly the same to that to fast acoustic forcing, we do not show the former for brevity.
For entropy and acoustic forcing, the magnitudes of  stagnation-point vorticity perturbation are comparable. {\color{cyan}Specifically, $\Omega_{\xi 0}$ under acoustic forcing is around 1.1 times greater than that under entropy forcing. This disparity is attributed to the greater freestream density perturbation for acoustic forcing, as demonstrated in (\ref{eq:estimate_FS}).} Notably, the response to acoustic forcing is slightly stronger. Furthermore, both responses increase monotonically with $k_3$.

For freestream vortical forcing, the magnitude of the stagnation response is significantly weaker than that for entropy or acoustic forcing. This is attributed to the absence of the freestream density perturbation $\hat \rho_\infty$ for vortical disturbances, rendering the freestream perturbation $O(M^{-1})$ smaller. Furthermore, as revealed by (\ref{eq:estimate_FS}), the post-shock perturbation under vortical forcing scales with   $\hat \Omega_2/k_3$. Therefore, for a prescribed $\hat\Omega_2=1$,   the stagnation response decreases with increase of $k_3$. This trend contrasts sharply with the behavior observed for acoustic and entropy forcing.

 In figure \ref{fig:asymptotic_entropy}(b), we further compare the composite solution profiles of $\tilde u_\eta$ and $\tilde w$ from the asymptotic model with SF-HLNS results for representative cases. The $\tilde w$ profiles  exhibit clear peaks within the boundary layer, with the peak magnitude $||\tilde w||_\infty$ increasing with  $k_3$. Again, close agreement is observed between the two methods, confirming the accuracy of our asymptotic analysis.

\subsection{Discussion}
 Based on asymptotic analysis of the double-layered structure near the centerline (regions III and IV), a simplified model is developed to determine the perturbation response at the stagnation point. The analysis clearly reveals the underlying mechanism responsible for the high-amplitude streamwise vorticity in the stagnation region: as the post-shock perturbation decelerates  to the stagnation point, streamwise vorticity is amplified to conserve momentum, a process termed the 'slow-down convection' mechanism.
Unlike in classical boundary layers, where the main flow convects along the body surface, this mechanism provides a general means of exciting high-amplitude vorticity perturbations at the leading edge of a blunt body. The vortical perturbation excited within the boundary layer acts as a seed for downstream streak formation, thereby constituting a key stage in the receptivity of non-modal streaks in hypersonic boundary layers with blunt leading edges.
Furthermore, the asymptotic model shows that for large Reynolds number $R$, the vorticity response scales as $R^{1/2}$. Since the Reynolds number is defined using the nose radius as the reference length, this implies that receptivity efficiency increases for larger bluntness, consistent with experimental observations that transition is promoted as bluntness increases in moderately to highly blunt configurations. A numerical justification of this $R^{1/2}$ scaling relation is also provided in Part III \cite{Sun2026}.

Quantitatively, for low-frequency non-modal receptivity, the magnitudes of the freestream perturbations $(\hat u_\infty,\hat \rho_\infty,\hat p_\infty)$ play the dominant role, with their magnitudes under unity energy norm   estimated in (\ref{eq:estimate_FS}). Clearly, the post-shock perturbations determined by the linearised R-H relation for freestream acoustic and entropy forcing are  much greater than that of vortical forcing. Correspondingly, the stagnation  perturbation vorticities excited by acoustic and entropy forcing are significantly stronger, rendering them more relevant for practical applications. This is somewhat counterintuitive:  introducing a freestream vortical perturbation actually results in less efficient excitation of streamwise perturbation vorticity, which is primarily caused by the distortion of the oncoming perturbations by the bow shock. Moreover, the asymptotic model reveals that the receptivity efficiency is identical for fast and slow acoustic waves and does not depend on the acoustic declination angle. For both acoustic and entropy forcing, the perturbation response at the stagnation point increases with the spanwise wavenumber $k_3$.

Since the asymptotic expansions employed  here rely on the small parameter $R^{-1/2}$, some numerical discrepancy between the asymptotic predictions and the SF‑HLNS results is understandable. However, this discrepancy decreases consistently as $R$ increases; {\color{red}a numerical evidence is shown in figure 6 of Part III of this work series}. Moreover, although the present theory is applied in the $M=O(1)$ regime,  the hypersonic asymptotic regime is expected to appear at sufficiently high Mach numbers. In that regime, both the governing equations and boundary conditions can be written in an  $M$-independent form, with  temperature scaled by $M^2$. {\color{red}Note that for very high temperatures, another viscosity law may be employed.}

\section{Evolution of non-modal streaks in the downstream region}
\label{sec:LPSE}
Our next step is to use the composite solution obtained from asymptotic analysis (\ref{eq:composite}) as an   initial perturbation for calculating the formation of non-modal streaks downstream. {\color{cyan}This corresponds to step (iv) outlined at the beginning of Section \ref{sec:asymptotic_theory}}.

{\color{cyan}\subsection{Scaling arguments for the non-modal-streak evolution}}

\begin{figure}
\begin{center}
  \includegraphics[width=0.95\textwidth]{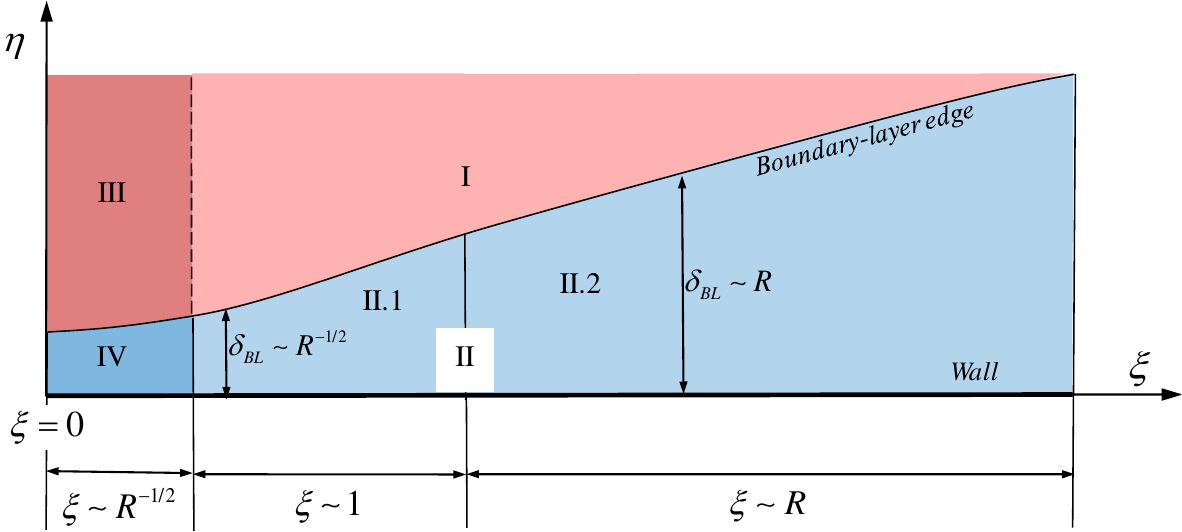}
  \caption{Sketch of the the distinct regions in the $\xi$-$\eta$ plane.}\label{fig:sketch4}
  \end{center}
\end{figure}

{\color{cyan}Figure \ref{fig:sketch4} displays the distinct flow regions in the $\xi$-$\eta$ plane, where   $\xi=0$ corresponds to the stagnation point. The boundary layer develops from the stagnation point to downstream. Unlike sharp-leading-edge configurations,  the boundary layer possesses a finite thickness at the stagnation point and grows proportionally to   $\xi^{-1/2}$ in region II. The boundary-layer structure is divided into region IV, occupying the $O(R^{-1/2})$ neighborhood of the stagnation point, and region II, extending from $\xi=O(1)$ to $\xi=O(R)$. We further decompose region II into subregions II.1 and II.2.

}

\begin{itemize}
    \item \textit{{\color{cyan}In region II.1, where $\xi\sim 1$, the boundary-layer thickness retains the same magnitude as that in region IV, implying  $\eta\sim R^{-1/2}$. Consequently, the corresponding magnitudes of the base flow are
    }}
\begin{equation}
    (\bar u_\xi,\bar\rho,\bar T,\bar P)\sim 1, \quad \bar u_\eta\sim R^{-1/2},
\end{equation}
and the perturbations,  which match the stagnation‑region solution (region IV), scale as
\begin{equation}
    (\tilde u_\xi,\tilde w,\tilde \rho,\tilde T,\tilde p)\sim 1,\quad \tilde u_\eta\sim R^{-1/2}.
\end{equation}

\item \textit{{\color{cyan}In region II.2, where $\xi\sim R\gg 1$, the boundary-layer thickness extends to $\eta=O(R^{-1/2}\sqrt{\xi})\sim O(1)$.} The base flow in this region displays similarity. Here, the boundary‑layer thickness becomes comparable to the spanwise wavelength $2\pi/k_3$ for $k_3=O(1)$. As indicated in ZD25, transient growth saturates at a location where the boundary-layer thickness matches the spanwise wavenumber. From the continuity equation, the scaling relations of the base flow read
}
\begin{equation}
        (\bar u_\xi,\bar\rho,\bar T,\bar P)\sim 1, \quad \bar u_\eta\sim R^{-1}.
\end{equation}
Given that the lateral perturbation velocity $\tilde w$ follows the same magnitude  as  region IV, and considering  the continuity equation, the magnitudes of the perturbation field can be estimated as
\begin{equation}
      (\tilde u_\xi, \tilde \rho,\tilde T)\sim R,\quad  (\tilde u_\eta,\tilde w,\tilde p)\sim 1.
\end{equation}
\end{itemize}
Therefore, as the perturbations evolve from the stagnation region (region IV) to the downstream plate, the perturbation streamwise velocity $\tilde u_\xi$ is amplified from $O(R^{-1/2})$ in region IV to $O(R)$ where $\xi \sim R$. This exemplifies the classical lift-up mechanism.

In the region where $\xi\sim 1$, the perturbations are governed by the linearised boundary-layer equation, with the base-flow non-parallelism playing a leading-order role. However, for $\xi\sim R$, where the base flow approaches the compressible Blasius solution, the  boundary-region equation  becomes a good approximation. This process aligns with the mechanism described in \cite{Leib1999} and \cite{Ricco2007response}.

{\color{cyan}Let us now revisit the BRE framework of \cite{Leib1999} and \cite{Ricco2007response} to draw a closer comparison with the present theory. In \cite{Leib1999}, the Reynolds number was defined based on the spanwise wavelength of the freestream forcing, which is assumed to be $O(1)$ in the present theory. Therefore, the Reynolds numbers in both studies are directly comparable. Furthermore, both theories assume the low-frequency feature of the non-modal perturbations, ensuring the formation of the downstream streaks due to the lift-up mechanism. The three regions identified in figure 1 of \cite{Leib1999} correspond to regions I, II.1, and II.2 in the present study, respectively. However, since \cite{Leib1999} excluded the blunt nose, the mechanism for initial perturbation formation differs intrinsically from that described herein. To recover the regime studied in \cite{Leib1999}, one must take the nose radius $r^*$ to zero,  implying that the Reynolds number based on nose radius,   $R=\rho_\infty^* U_\infty^* r^*/\mu_\infty^*$, becomes vanishingly small. Given that the perturbation streamwise vorticity  $\Omega_{\xi 0}$ scales with $R^{1/2}$,  decreasing $R$ leads to a a significant reduction in the stagnation perturbation strength. This suggests that in the sharp-leading-edge limit, the 'slow-down convection' mechanism disappears due to the absence of the stagnation region.

Moreover, \cite{Leib1999} and \cite{Ricco2007response} focused exclusively on freestream vortical forcing, owing to its shared dispersion relation with non-modal perturbations. However, in hypersonic configurations, the presence of a bow shock fundamentally alters the receptivity process. The shock ensures that any incident perturbation, whether acoustic, entropy, or vortical, can excite all three perturbation types in the post-shock region. As indicated by (\ref{eq:estimate_FS}), the R-H relations imply that freestream acoustic and entropy perturbations are more efficient at exciting stagnation perturbations compared to vortical disturbances.

In conclusion, the present theory covers the BRE framework developed by \citep{Leib1999,Ricco2007response} for describing non-modal streak formation. However, the inclusion of blunt-nose configurations reveals a new physical mechanism, the 'slow-down convection' mechanism. This mechanism facilitates a more efficient excitation of boundary-layer perturbations near the stagnation point, providing the initial condition for subsequent region-II calculations. Furthermore, the present framework conveniently accounts for the detached bow shock, demonstrating that the boundary-layer response is significantly more receptive to freestream acoustic and entropy forcing compared to vortical forcing.
  }

\subsection{Numerical method to predict the  formation of the downstream streaks }
Overall, the perturbation evolution downstream of region IV is fully parabolic. {\color{cyan}From a rigorous asymptotic standpoint, the boundary-layer equation and the boundary-region equation should be employed to describe the perturbation evolution in subregions II.1 and II.2, respectively.} {\color{red}However, to achieve a unified numerical treatment valid for both subregions, we employ a finite-$R$ LPSE approach. This method retains most terms of the linearized N-S system, except for the elliptic terms, which are neglected based on the scaling analyses inherent to boundary-layer and boundary-region equations. The principal advantage of the LPSE method is that, through the transition from subregion II.1 to II.2, where the dominant balance shifts from boundary-layer to boundary-region dynamics, it ensures a smooth development at finite Reynolds numbers.}

The LPSE approach adopted in this study follows the methodology established in previous studies \citep{zhao2016improved,song2023effect,song2024influence}. First, a computational domain downstream of the stagnation region is selected, defined as $\xi\in[X_0,\xi_N]$ and $\eta\in[0,\eta_N]$, where $X_0$ is taken immediately downstream of the elliptic region, $\xi_N$ is positioned sufficiently far downstream, and $\eta_N$ lies within the inviscid region but below the shock location. Substituting the decomposition (\ref{eq:decompositioni}) into the linearised N-S equations and neglecting the $\partial_{\xi\xi}$ and $\partial_{\xi\eta}$ terms yields the governing equations
\begin{equation}
 \partial_\xi\tilde\phi=  {\cal L}_{LPSE} \tilde \phi,
\end{equation}
where ${\cal L}_{LPSE}$ represents a linear operator including  $\partial_\eta$ and $\partial_{\eta\eta}$ terms; its detailed expression is provided in \cite{ren2014competition} and \cite{zhao2016improved}. The $\partial_\xi$ term is discretized using a second-order backward difference scheme. The initial perturbation at $\xi=X_0$   is prescribed by the composite solution in (\ref{eq:composite}), allowing the downstream perturbation to be solved via a marching scheme. No-slip and adiabatic conditions are imposed at the wall, while damping conditions are applied at the upper boundary $\eta=\eta_N$. The latter is justified because, as shown in {figure} \ref{fig:baseflow}-(d), the non-modal receptivity is only excited within a narrow region near the centerline. {\color{cyan}Note that the LPSE method accommodates  non-zero frequencies, and  streak formation  driven by the lift-up mechanism is well captured provided that $\omega\sim R^{-1}$. However, since the maximum amplification occurs at zero frequency, the subsequent analysis focuses exclusively on the stationary case ($\omega=0$).}

\subsection{Numerical results of LPSE}
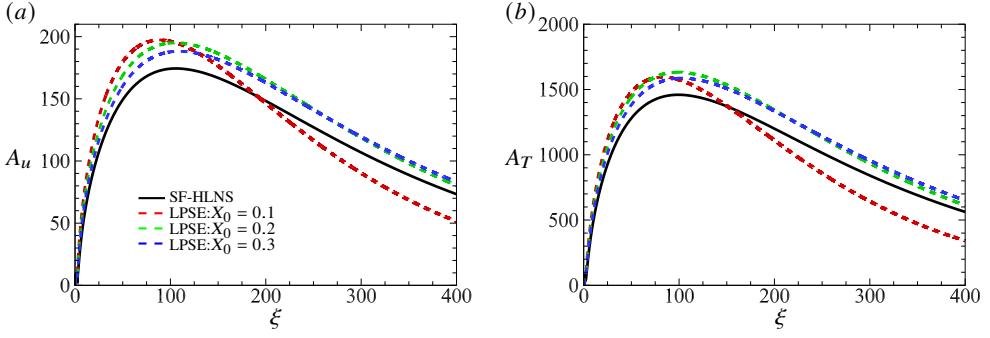
\begin{figure}
  \begin{center}
  \includegraphics[width=\textwidth]{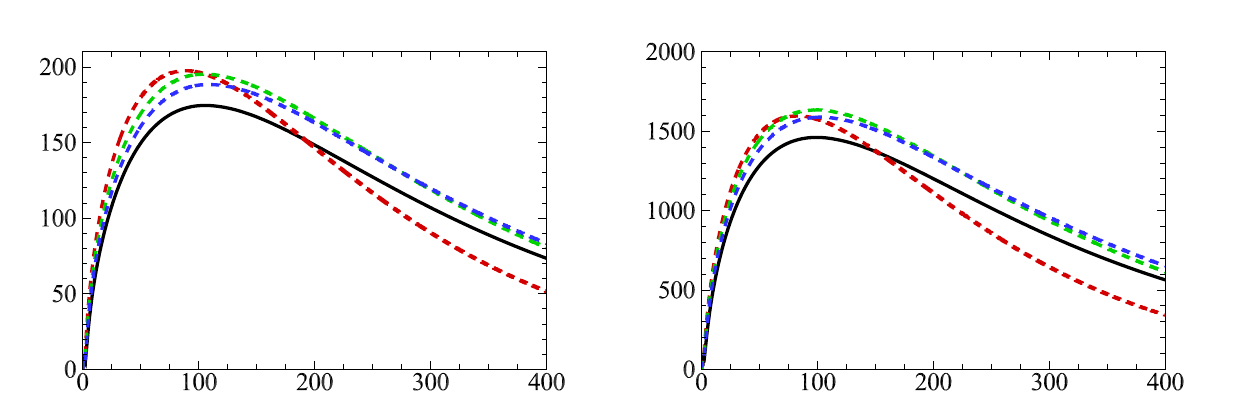}
  \put(-385,115){$(a)$}
  \put(-197,115){$(b)$}
  \put(-385,60){$A_u$}
  \put(-197,60){$A_T$}
  \put(-285,0){$\xi$}
  \put(-95,0){$\xi$}
  \put(-335,30) {\begin{tikzpicture}
    \draw[black,thick]   (0,0) -- (0.35,0);
    \draw[red,thick,dashed]   (0,-0.2) -- (0.35,-0.2);
    \draw[green,thick,dashed] (0,-0.4) -- (0.35,-0.4);
    \draw[blue,thick,dashed]  (0,-0.6) -- (0.35,-0.6);
    \end{tikzpicture}}
  \put(-323,46) {\fontsize{6pt}{6pt}\selectfont SF-HLNS}
  \put(-323,40) {\fontsize{6pt}{6pt}\selectfont LPSE:$X_0=0.1$}
  \put(-323,34) {\fontsize{6pt}{6pt}\selectfont LPSE:$X_0=0.2$}
  \put(-323,28) {\fontsize{6pt}{6pt}\selectfont LPSE:$X_0=0.3$}
  \caption{Streamwise evolution of perturbation amplitudes $A_u$ $(a)$ and $A_T$ $(b)$ for freestream entropy forcing with $\omega=0$, {$\vartheta= 0^{\circ}$} and $k_3=8$, obtained by LPSE with various initial positions  $X_0$. The SF-HLNS results (the same as the red line in figure \ref{fig:baseflow}-(d)) are plotted by solid lines for reference.}
  \label{fig:amplitude_sfhlns_lpse}
  \end{center}
\end{figure}

For an adiabatic base flow with $M=5.96$, $T_\infty^*=87K$, and a freestream entropy wave characterised by $\omega=0$, {$\vartheta= 0^{\circ}$} and $k_3=8$, we perform the LPSE calculations initialised at a position $\xi=X_0$. 
Figure \ref{fig:amplitude_sfhlns_lpse} presents several LPSE calculations initialised at different downstream positions $X_0$ relative to the stagnation point. For a very small $X_0$=0.1, the downstream evolution of the excited non-modal perturbation deviates significantly from the baseline SF-HLNS curve, with the discrepancy increasing with $\xi$. This indicates that $\xi=0.1$ lies within the elliptic region (region IV), where parabolised approximation cannot fully  capture the local perturbation physics. When the initial position is shifted slightly downstream to $X_0$=0.2 or 0.3, the downstream evolution of the non-modal perturbation closely matches the SF-HLNS prediction. {\color{red}A residual discrepancy persists, which can be attributed to the finite-$R$ effect neglected in the asymptotic model}. This agreement confirms that these starting locations  are outside the elliptic zone and governed by the parabolic regime. These observations are fully consistent with the SF-HLNS calculations obtained under forcing localised to a narrow region, as displayed in {figure} \ref{fig:baseflow}-(d).

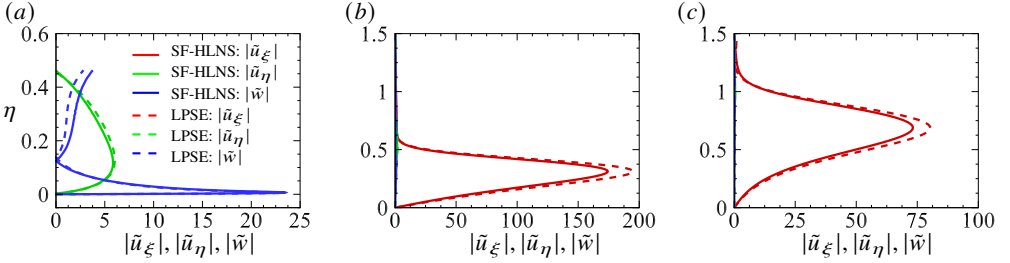
\begin{figure}
  \begin{center}
  \includegraphics[width=\textwidth]{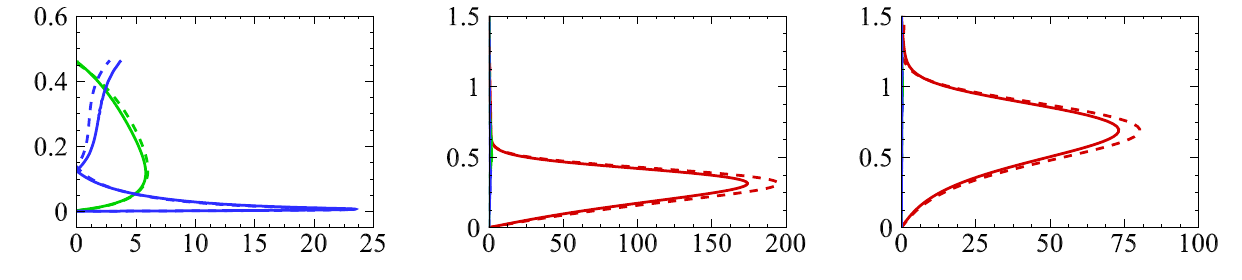}
  \put(-380,87){$(a)$}
  \put(-252,87){$(b)$}
  \put(-127,87){$(c)$}
  \put(-380,50){$\eta$}
  \put(-335,0){$|\tilde u_{\xi}|, |\tilde u_{\eta}|, |\tilde w|$}
  \put(-205,0){${|\tilde u_{\xi}|, |\tilde u_{\eta}|}, |\tilde w|$}
  \put(-80,0) {${|\tilde u_{\xi}|, |\tilde u_{\eta}|}, |\tilde w|$}
  \put(-332,36) {\begin{tikzpicture}
    \draw[red,thick]   (0,0) -- (0.35,0);
    \draw[green,thick] (0,-0.25) -- (0.35,-0.25);
    \draw[blue,thick]  (0,-0.5) -- (0.35,-0.5);
    \draw[red,thick,dashed]  (0,-0.8) -- (0.35,-0.8);
    \draw[green,thick,dashed] (0,-1.05) -- (0.35,-1.05);
    \draw[blue,thick,dashed]  (0,-1.3) -- (0.35,-1.3);
    \end{tikzpicture}}
  \put(-317,73) {\fontsize{6pt}{6pt}\selectfont SF-HLNS: $|\tilde u_{\xi}|$}
  \put(-317,65) {\fontsize{6pt}{6pt}\selectfont SF-HLNS: $|\tilde u_{\eta}|$}
  \put(-317,57) {\fontsize{6pt}{6pt}\selectfont SF-HLNS: $|\tilde w|$}
  \put(-317,49) {\fontsize{6pt}{6pt}\selectfont LPSE: $|\tilde u_{\xi}|$}
  \put(-317,41) {\fontsize{6pt}{6pt}\selectfont LPSE: $|\tilde u_{\eta}|$}
  \put(-317,33) {\fontsize{6pt}{6pt}\selectfont LPSE: $|\tilde w|$}
  \caption{Comparison of perturbation velocity profiles obtained by SF-HLNS (solid lines) and LPSE with $X_0=0.2$ (dashed lines). (a): $\xi=0.2$; (b): $\xi=100$; (c): $\xi=400$.}
  \label{fig:eigenf_sfhlns_lpse}
  \end{center}
\end{figure}

\begin{figure}
  \begin{center}
  \includegraphics[width=0.48\textwidth]{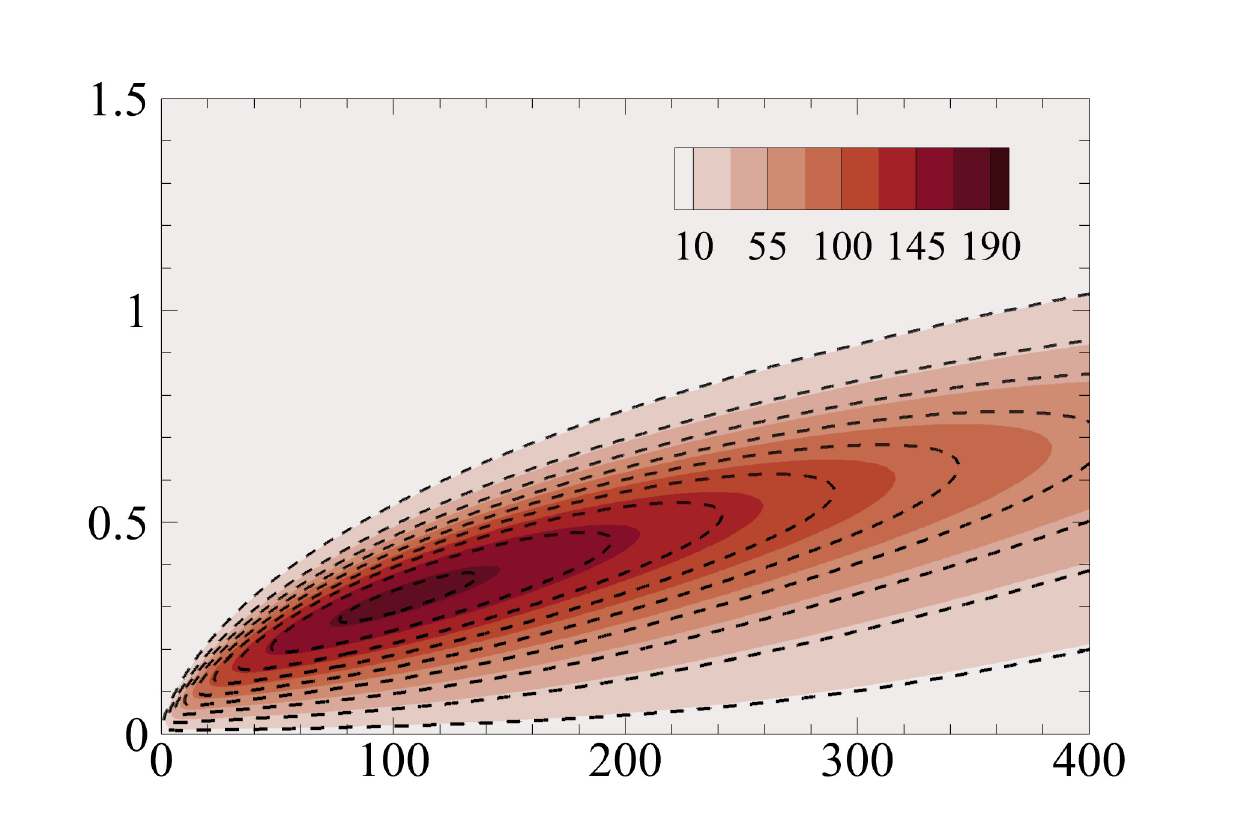}
  \includegraphics[width=0.48\textwidth]{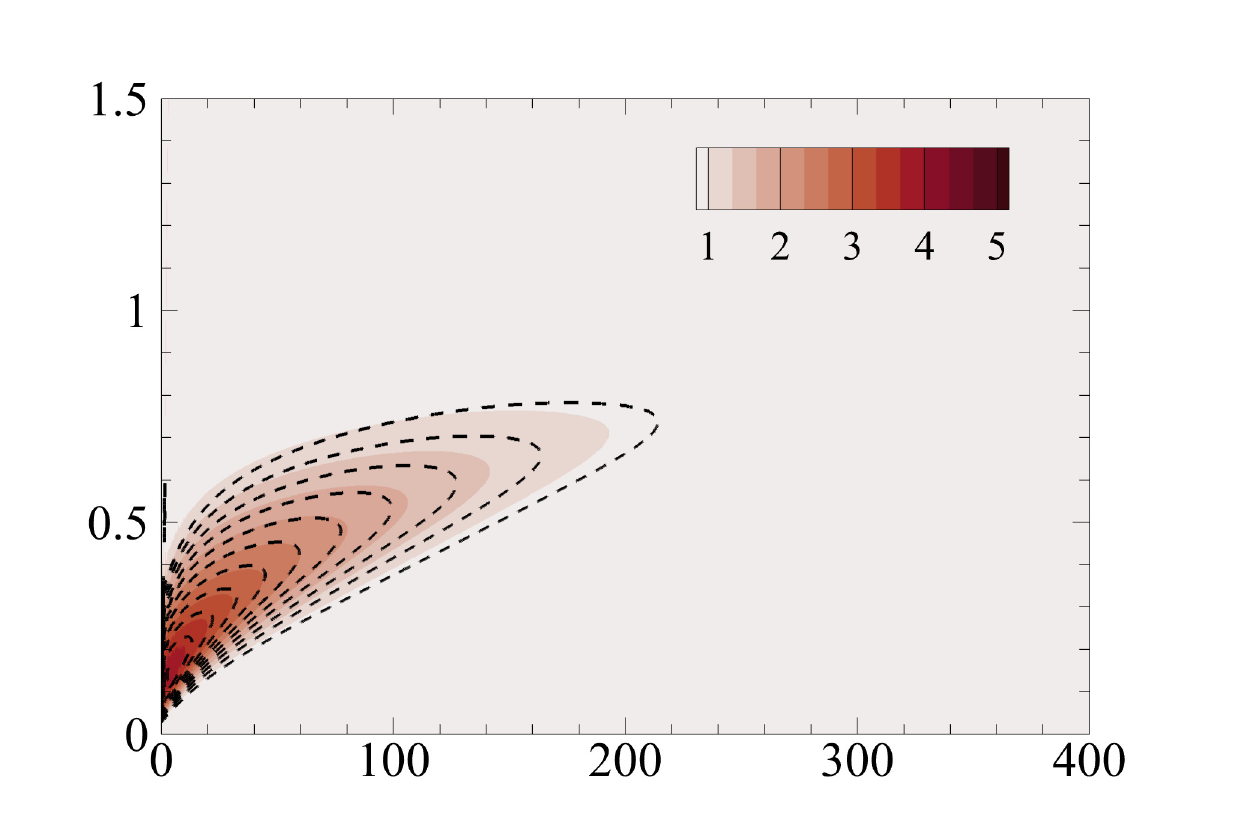}
  \put(-380,115){$(a)$}
  \put(-192,115){$(b)$}
  \put(-380,60){$\eta$}
  \put(-280,0){$\xi$}
  \put(-95,0){$\xi$}
  \put(-290,92){$|\tilde u_{\xi}|$}
  \put(-103,92){$|\tilde u_{\eta}|$}\\
  \includegraphics[width=0.48\textwidth]{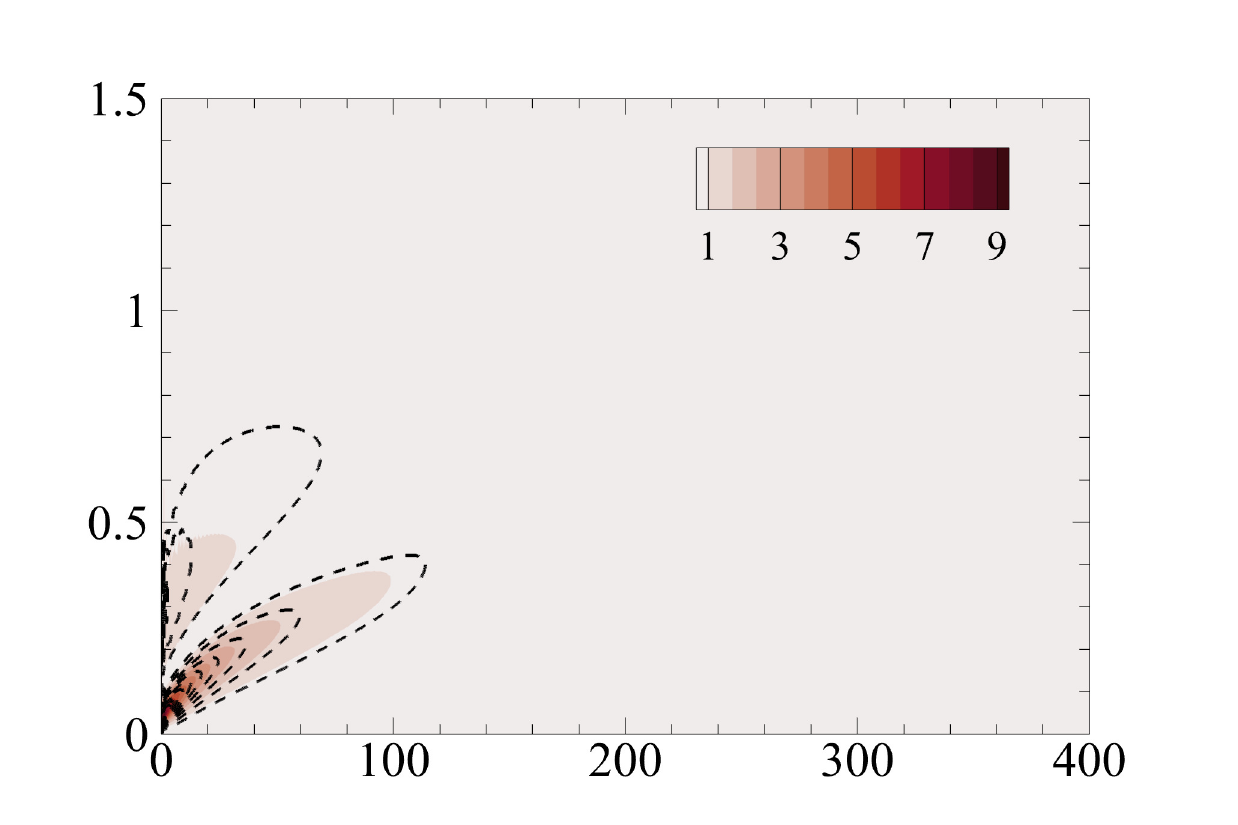}
  \put(-192,115){$(c)$}
  \put(-192,60){$\eta$}
  \put(-95,0){$\xi$}
  \put(-100,92){$|\tilde w|$}
  \caption{{\color{magenta}Comparison of the perturbation velocity contours  obtained by SF-HLNS calculations (flood) and LPSE (dashed lines) with $X_0=0.2$.}}
  \label{fig:cont_eigenf_sfhlns_lpse}
  \end{center}
\end{figure}
In figure \ref{fig:eigenf_sfhlns_lpse}, we  examine the perturbation  profiles at different streamwise positions based on $X_0=0.2$. In the nose region,  the perturbation velocity components {$\tilde u_{\eta}$} and $\tilde w$  dominate, indicating a roll structure. In the downstream region, as shown in panels (b) and (c), the perturbation streamwise velocity {$\tilde u_{\xi}$} attains a much greater magnitude than {$\tilde u_{\eta}$} and $\tilde w$, reflecting a characteristic streak structure. {\color{magenta}
This feature is further illustrated in figure \ref{fig:cont_eigenf_sfhlns_lpse}, which clearly demonstrates the decay of $\tilde u_\eta$ and $\tilde w$ and the transient growth of $\hat u_\xi$.}
This is a typical demonstration of the classical lift-up mechanism.
Notably, the perturbations obtained from LPSE calculations agree closely with those from SF‑HLNS computations, confirming the accuracy of the simplified approach.

Thus,  the combination of the composite solution from asymptotic analysis and downstream LPSE calculation offers a reduced model for rapid prediction of non‑modal receptivity. In practice, the base flow must first be computed by solving the steady N-S equations. In particular, the second streamwise derivative of pressure at the centerline, $\partial_\xi^2\bar P(\xi=0,\eta)$, is needed for asymptotic analysis. Because the pressure exhibits zero wall‑normal gradient within the boundary layer to leading order, the steady N-S solution for $\partial_\xi^2\bar P(\xi=0,\eta)$ is  essentially identical to the inviscid solution obtained from the region‑I Euler equations. Hence,   $\partial_\xi^2\bar P(\xi=0,\eta)$  is taken as $R^{-1}P_2(\eta)$  defined in (\ref{eq:base_flow_quantities}), which serves as  the sole input for deriving the composite solution. Using the steady N-S solution as the base flow and the composite solution as the initial perturbation, LPSE is then applied to compute the linear evolution of non‑modal perturbations downstream. This model is based on the physical mechanism identified by the asymptotic analysis, which captures the dominant factors governing the receptivity process, and its accuracy improves with increasing Reynolds number.

\section{Concluding remarks}
\label{sec:conclusion}
This series of studies focuses on a canonical problem: the receptivity of non‑modal perturbations in hypersonic boundary layers over blunt bodies, an issue of practical importance for predicting bypass transition. Due to the presence of a detached bow shock, quantifying how freestream disturbances interact with the shock and eventually excite downstream non‑modal perturbations has long been challenging. The SF‑HLNS framework developed in Part I of this series provides an efficient and accurate means to quantify receptivity efficiency under a wide range of freestream forcing and control parameters.
In this paper, we perform high‑Reynolds‑number asymptotic analysis to uncover the underlying receptivity mechanism.

SF‑HLNS calculations (figure \ref{fig:baseflow}-(d)) show that receptivity is most sensitive to forcing imposed in a {localised} region near the centerline. This observation motivates a focused examination of the  $O(R^{-1/2})$ neighbourhood around the centreline. Two distinct layers emerge in this region: region III, which contains the inviscid flow from the shock toward the wall, and region IV, which lies within the $O(R^{-1/2})$ vicinity of the wall. Further downstream, we  recover the classical inviscid layer (region I) and the viscous boundary layer (region II).
\begin{figure}
  \begin{center}
  \includegraphics[width=\textwidth]{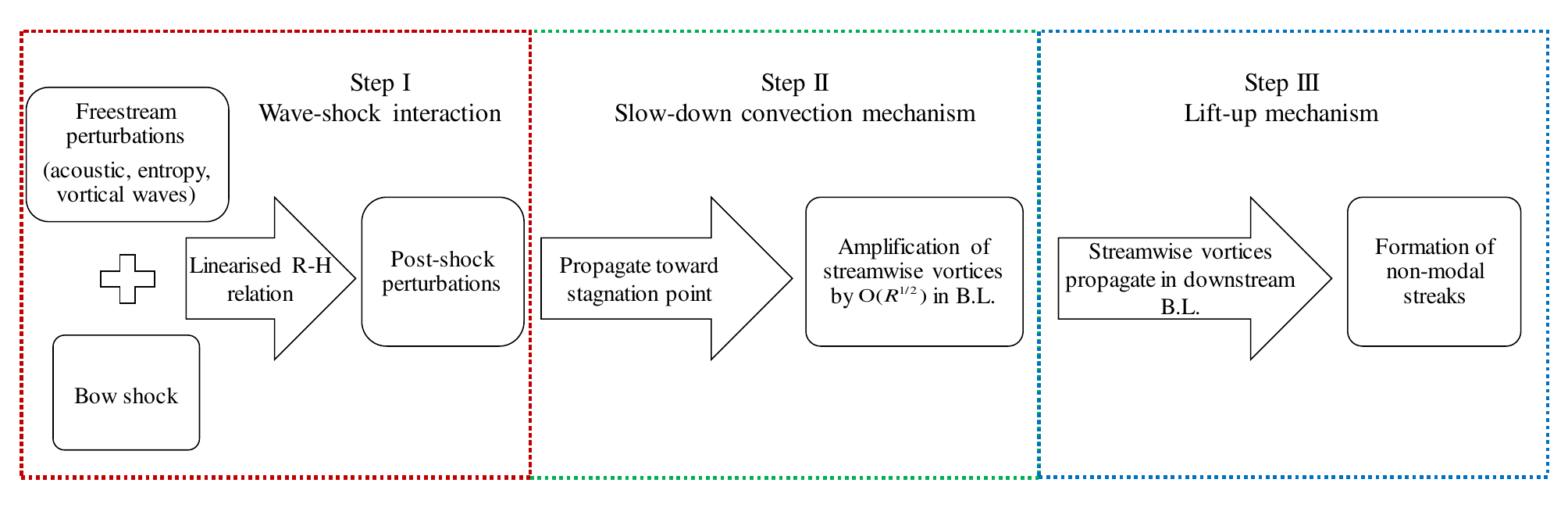}
  \caption{Schematic of the physical mechanism for non-modal receptivity.}
  \label{fig:mechanism}
  \end{center}
\end{figure}

The physical mechanism of non‑modal receptivity can be summarized in three steps, as sketched in figure \ref{fig:mechanism}. 
\begin{itemize}[leftmargin=2em]
\item \ \ First, freestream perturbations, whether acoustic, entropy, or vortical waves, interact with the bow shock around the centreline, exciting post-shock perturbations that serve as the upper boundary condition for   region III. This interaction, governed by the linearised R-H relation, involves shock oscillation and renders freestream acousitc and entropy forcing more effective.  

\item \ \ Second, these post-shock perturbations propagate toward the stagnation point. The streamwise vorticity amplifies due to the deceleration of the base flow, a process termed the slow-down convection mechanism, which contrasts with classical boundary-layer behavior. Amplification ceases upon entering the viscous boundary layer (B.L. in Figure \ref{fig:mechanism}), reaching an $O(R^{1/2})$ factor. 

\item \ \  Third, the excited streamwise vortices propagate downstream within the boundary layer,  generating longitudinal streaks (non-modal perturbations) via the lift-up mechanism.
\end{itemize}

It can be readily explained by considering the shock-perturbation interaction (\ref{eq:estimate_FS}) that the non-modal receptivity to low-frequency freestream fast  and slow acoustic waves is identical to leading order and independent of the declination angle. The receptivity efficiency for acoustic forcing is comparable to that for freestream entropy forcing. In  these cases, the stagnation‑point response of the perturbation streamwise vorticity increases with  $k_3$ and  exhibits a phase lag of $\pi$ relative to the freestream forcing. In contrast, the stagnation response to freestream vortical forcing is much weaker. The excited perturbation streamwise vortices at the stagnation region serve as the direct seed for downstream non‑modal perturbation formation, with their intensities directly linked.

Building on these physical observations, we construct a reduced model for rapid estimation of receptivity efficiency. The first step is to compute the composite solution (\ref{eq:composite}) in the stagnation region using the asymptotic analysis. The second step is to perform LPSE calculations starting immediately downstream of the elliptic stagnation region, using the composite solution as the initial perturbation. Comparisons with SF‑HLNS results confirm the accuracy of both the initial composite solution and the downstream evolution of the non‑modal perturbations.

{\color{cyan}The second step of the reduced model is equivalent to the BRE framework as in \cite{Leib1999} and \cite{Ricco2007response}. However, the first step specifically accounts for the blunt-nose effect, uncovering a distinct 'slow-down convection' mechanism that differentiates it from the conventional BRE framework. This mechanism facilitates a more efficient excitation of perturbations near the stagnation point, providing the initial condition required for the second step. Furthermore, by integrating the effect of the detached bow shock, our reduced model demonstrates that the boundary layer is significantly more receptive to freestream acoustic and entropy disturbances than to vortical ones, a phenomenon that lies beyond the descriptive scope of the conventional BRE framework.}

It should be noted that the present analysis is based on hypersonic boundary layers with adiabatic walls, for which the thermal boundary layer does not develop in the stagnation region.
Because of its simplicity, this configuration best illustrates the essential receptivity mechanism. While cold‑wall configurations are common in practical applications, their influence on receptivity will be examined in Part III of this series \citep{Sun2026}.

 \vspace{.4cm}
 \noindent\textbf{Funding.} {This work is supported by National Natural Science Foundation of China (grant nos. 92371104, 12588201, 12372222),   the CAS Strategic Priority Research Program (no. XDB0620102) and CAS project for Young Scientists in Basic Research (YSBR-087).}

  \vspace{.4cm}
  \noindent\textbf{Declaration of interests.}
  {The authors report no conflict of interest.}

\bibliographystyle{jfm}

\bibliography{receptivity_mechanism3_R1}

\end{document}